\documentclass[12pt]{article}

\usepackage[utf8]{inputenc}
\usepackage{cancel}
\usepackage{amssymb}
\usepackage{soul}
\usepackage{tabularx}
\usepackage{xcolor}
\usepackage{booktabs}
\usepackage{subcaption} % Added by NN
\usepackage{colortbl}
\usepackage{authblk}
\DeclareUnicodeCharacter{00A0}{ }

\newcolumntype{C}{>{\centering\arraybackslash}X}
\usepackage[T1]{fontenc}
\usepackage{epsfig}
\usepackage{latexsym}
\usepackage{graphicx}
\usepackage{amsmath}
\usepackage{amsfonts}   % if you want the fonts
\usepackage{amssymb}    % if you want extra symbols
\usepackage{float}
\usepackage{bm}
\usepackage{url}
\usepackage{hyperref} 
\usepackage[nodisplayskipstretch]{setspace}
\def\lsim{\raise0.3ex\hbox{$\;<$\kern-0.75em\raise-1.1ex\hbox{$\sim\;$}}}
\def\gsim{\raise0.3ex\hbox{$\;>$\kern-0.75em\raise-1.1ex\hbox{$\sim\;$}}}
\def    \beq            {\begin{equation}}
\def    \eeq            {\end{equation}}
\def    \bea           {\begin{eqnarray}}
\def    \eea           {\end{eqnarray}}

\def \mn{\mu\nu{\rm SSM}}

\def\g2{{\rm GeV}^2}
\def\sw2{sin^2 \theta_w}

\def\a^tau{\alpha_{\tau}}

\def\beq{\begin{equation}}
\def\eeq{\end{equation}}
\def\beqa{\begin{eqnarray}}
\def\eeqa{\end{eqnarray}}

\newcommand{\newc}{\newcommand}
\newc\BR{BR}
\newc{\akappa}{A_{\kappa} }
\newc\deltagmtwo{\delta (g-2)_{\mu}} 
\newc\deltaamu{\Delta a_{\mu}}

\def\anti{\overline}

\def\la{\lambda}

\def\ka{\kappa}

\newc{\haa}{BR\(h_1\to a_1 a_1\)}
\newc{\abb}{BR\(a_1\to b\anti{b}\)}
\newc{\hbb}{BR\(h_1\to b\anti{b}\)}
\newc{\abund}{\Omega h^2}
\newc\bsgamma{b\rightarrow s \gamma }
\newc\bxsgamma{\overline{B}\rightarrow X_{s}\gamma}
\newc\brbsgamma{\BR(\overline{B}\rightarrow X_s\gamma)}

\allowdisplaybreaks

\usepackage{array, makecell}
\usepackage{boldline}
\usepackage[nosort]{cite}

\title{\bf{
Sampling the $\mu\nu$SSM for displaced decays of the tau left sneutrino LSP at the LHC
}}
\author[a,b]{Essodjolo Kpatcha\thanks{kpatcha.essodjolo@uam.es}}
\author[a,b,h]{I\~naki Lara\thanks{inaki.lara@csic.es}}
\author[c,d]{Daniel~E.~L\'opez-Fogliani\thanks{daniel.lopez@df.uba.ar}}
\author[a,b]{Carlos~Mu\~noz\thanks{c.munoz@uam.es}} 
\author [e]{Natsumi Nagata\thanks{natsumi@hep-th.phys.s.u-tokyo.ac.jp}}
\author [f]{Hidetoshi Otono\thanks{otono@phys.kyushu-u.ac.jp}}
\author[g]{Roberto~Ruiz~de~Austri\thanks{rruiz@ific.uv.es}}

\affil[a]{Departamento de F\'{\i}sica Te\'{o}rica, Universidad Aut\'{o}noma de Madrid (UAM),
Campus de Cantoblanco, 28049 Madrid, Spain}
\affil[b]{Instituto de F\'{\i}sica Te\'{o}rica (IFT) UAM-CSIC, 
  Campus de Cantoblanco, 28049 Madrid, Spain}
  \affil[c]{Instituto de F\'isica de Buenos Aires UBA \& CONICET, Departamento de F\'isica,
 Facultad de Ciencia Exactas y Naturales, Universidad de Buenos Aires, 
1428 Buenos Aires, Argentina}
\affil[d]{
{Pontificia Universidad Cat\'olica Argentina, 
1107 Buenos Aires, Argentina}}
\affil[e] {Department of Physics, University of Tokyo, 
Tokyo 113-0033, Japan}
\affil[f] {Research Center for Advanced Particle Physics, Kyushu University, 
Fukuoka 819-0395, Japan}
  \affil[g]{Instituto de F\'{\i}sica Corpuscular CSIC--UV, 
c/Catedr\'atico Jos\'e Beltr\'an 2, 46980 Paterna,
 Valencia, Spain}
\affil[h] {Center for Theoretical Physics of the Universe, Institute for Basic Science (IBS),
Daejeon, 34126, Korea}

\date{}

\begin{document}

\maketitle

\begin{abstract}

Within the framework of the $\mn$, 
a displaced dilepton signal is expected at the LHC from 
the decay of a tau left sneutrino as the lightest supersymmetric
particle (LSP) with a mass in the range $45 - 100$ GeV.
We compare the predictions of this scenario  
with the ATLAS search
for long-lived particles using 
displaced lepton pairs in $pp$ collisions,
considering an optimization of the trigger requirements by means of a high level trigger that exploits tracker
information.
The analysis is carried out in the general case of three families of
right-handed neutrino superfields, where all the neutrinos get contributions to their masses at tree level.
To analyze the parameter space,
we sample the $\mu\nu$SSM for a tau left sneutrino LSP with proper decay length $c\tau > 0.1$ mm
using a likelihood data-driven method, and paying special attention to reproduce the current experimental data on neutrino and Higgs physics, as well as flavor observables.
The sneutrino is special in the $\mn$ since its couplings have to be chosen so that the
neutrino oscillation data are reproduced.
We find that important regions of the parameter space can be probed 
at the LHC run 3.

\end{abstract}


%%%%%%%%%%%%%%%%%%%%%%%%%%%%%%%%%%%%%%%%%%%%%%%%%%%%%%%%%%%%%%%%%%
\tableofcontents 
%\listoffigures
%\listoftables
%%%%%%%%%%%%%%%%%%%%%%%%%%%%%%%%%%%%%%%%%%%%%%%%%%%%%%%%%%%%%%%%%%

\section{Introduction}
\label{section1}

The search for low-energy supersymmetry (SUSY) is one of the main goals of the LHC. This search
has been focused mainly on signals with missing transverse energy (MET) inspired in $R$-parity conserving (RPC) models, such as the
minimal supersymmetric standard model (MSSM) \cite{Nilles:1983ge,Haber:1984rc,Martin:1997ns}. There, significant bounds on sparticle masses have been obtained~\cite{Tanabashi:2018oca}, especially for strongly interacting sparticles whose masses must be above about 1 TeV \cite{Aaboud:2017vwy, Sirunyan:2017kqq}.
Less stringent bounds of about 100 GeV have been obtained for weakly interacting sparticles, and even the bino-like neutralino is basically not constrained due to its small pair production cross section.
Qualitatively similar results have also been obtained in the analysis of simplified 
$R$-parity violating (RPV) scenarios with trilinear lepton- or baryon-number 
violating 
terms~\cite{Barbier:2004ez}, assuming a single channel available for the decay of the LSP into leptons.
However, this assumption is not possible in other RPV scenarios, such as 
the `$\mu$ from $\nu$' supersymmetric standard 
model ($\mn$)~\cite{LopezFogliani:2005yw},
where the several decay branching ratios (BRs) of the LSP significantly decrease the signal.
This implies that the extrapolation of the usual bounds on sparticle masses to the $\mn$ is not applicable.

The most recent analyses of signals at the LHC for LSP candidates in the
$\mn$ have been dedicated 
to the left sneutrino~\cite{Ghosh:2017yeh,Lara:2018rwv}, and to the bino-like neutralino~\cite{Lara:2018zvf}.\footnote{The phenomenology of a neutralino LSP was analyzed in the past 
in Refs.~\cite{Ghosh:2008yh,Bartl:2009an,Ghosh:2012pq,Ghosh:2014ida}.
{In the recent works~\cite{Biekotter:2017xmf,Biekotter:2019gtq}, in addition to perform the complete one-loop renormalization of the neutral scalar sector of the $\mn$, interesting scenarios with right sneutrinos lighter than the standard model-like Higgs boson were studied.}}
In the latter case, it was shown that no points of the parameter space of the $\mn$ were excluded when the left sneutrino is the next-to-LSP (NLSP) and hence a suitable source of binos.
In the region of bino (sneutrino) masses $110-120$ ($120-140$) GeV it was found a tri-lepton signal compatible with the local excess reported by ATLAS~\cite{Aaboud:2018sua}.
If this excess were due to a statistical fluctuation,\footnote{{The recent emulated recursive jigsaw reconstruction~\cite{ATLAS-CONF-2019-020} confirmed the 3$\sigma$ excess with 36 fb$^{-1}$, but sees only a small 1.27$\sigma$ excess of data with respect to predictions with full 139 fb$^{-1}$.}} the 
prospects for the bounds on the parameter space of the sneutrino-bino mass in the $\mn$ 
were discussed for the 13-TeV search with an integrated luminosity of 100 and 300 fb$^{-1}$.

{Concerning the left sneutrino LSP, 
in Ref.~\cite{Ghosh:2017yeh} the prospects for detection of signals
with di-photon plus leptons or MET from neutrinos, and multi-leptons, 
from the pair production of left sneutrinos/sleptons and their prompt decays ($c\tau\lsim$ 0.1 mm), were analyzed.
A significant evidence is expected only in the mass range of about 100 to 300 GeV. 
 The mass range of 45 to 100 GeV (with the lower limit imposed not to disturb the decay width of the $Z$) was covered in Ref~\cite{Lara:2018rwv}
 for the tau left sneutrino ($\widetilde{\nu}_{\tau}$) LSP.
 First, it was checked that no constraint on the $\widetilde{\nu}_{\tau}$ mass is obtained from previous searches.
 In particular, since the sneutrino has several relevant decay modes, the LEP lower bound on its mass mass of about 90 GeV~\cite{Abreu:1999qz,Abreu:2000pi,Achard:2001ek,Heister:2002jc,Abbiendi:2003rn,Abdallah:2003xc}
obtained under the assumption of BR one to leptons, via trilinear RPV couplings, is not applicable.
Similar conclusions were obtained from LEP mono-photon search (gamma+MET)~\cite{Abdallah:2003np}, and LHC mono-photon and mono-jet (jet+MET) searches~\cite{Aaboud:2016uro,Aaboud:2016tnv}.
Concerning LEP searches for staus~\cite{Abreu:1999qz,Abreu:2000pi,Achard:2001ek,Heister:2002jc,Abbiendi:2003rn,Abdallah:2003xc}, in the $\mn$ the left stau does not decay directly but through an off-shell $W$ and a $\widetilde{\nu}_{\tau}$, and therefore searches for its direct decay are not relevant in this model. Although the sneutrino mass can in principle be
constrained using searches for final states as those of the $\mn$
from the production of a pair of $\widetilde{\nu}_{\tau}$ from staus, it
was also checked in Ref~\cite{Lara:2018rwv} that this is not the case. 
Then, the displaced-vertex decays of the $\widetilde{\nu}_{\tau}$
LSP producing signals with di-lepton pairs was studied.
Using the present data set of the ATLAS 8-TeV dilepton search~\cite{Aad:2015rba}, the conclusion was that
one can constrain the sneutrino in some regions of the parameter space of the $\mn$, especially when the Yukawa couplings and mass scale of neutrinos are rather small. In order to improve the sensitivity of this search, it was proposed %in~\cite{Lara:2018rwv}
an optimization of the trigger requirements exploited in ATLAS based on a high level trigger that utilizes the tracker information.}

The above analyses were carried out in the simplest case of the $\mn$ with one right-handed neutrino superfield. Thus only one of the light neutrinos gets a nonvanishing tree-level contribution to its mass, whereas the other two masses rely on loop corrections. Basically, the only experimental constraint imposed
in those works was that
the heavier neutrino mass should be in
the 
range $m_\nu \sim$ [0.05, 0.23] eV, i.e. below the upper bound on the sum of neutrino masses $\sim 0.23$ eV \cite{Ade:2015xua}, and above the square root of the mass-squared difference 
$\Delta m_{\rm atm}^2\sim 2.42\times 10^{-3}\mathrm{eV}^2$ \cite{An:2015rpe}.
{In addition, the simplified assumption that all neutrino Yukawas have the same value was also applied.}
Although these analyses were useful to get a first idea of the accelerator constraints on the left sneutrino LSP, the lack of experimental bounds on the masses of the superpartners in the $\mn$ makes it peremptory a detailed study reproducing the whole neutrino physics.
This is the aim of this work. 
We will reconsider the analysis of Ref.~\cite{Lara:2018rwv}, but in the
context of the $\mu\nu$SSM with three families of right-handed neutrino superfields where all the neutrinos get contributions to their masses at tree level, {and different values of the neutrino Yukawas are necessary to reproduce neutrino physics.}
In particular, we will study the constraints on the parameter space by sampling the model to get
the $\widetilde{\nu}_{\tau}$ LSP in the range of masses $45-100$ GeV, with a decay length of the order of the millimeter. We will pay special attention
to reproduce the experimental
neutrino masses and mixing angles~\cite{Capozzi:2017ipn,deSalas:2017kay,deSalas:2018bym,Esteban:2018azc}. 
The different values of the neutrino Yukawas will imply that certain regions of the parameter space are excluded by the LEP analysis, unlike the result of Ref~\cite{Lara:2018rwv}.
In addition, we will impose on the resulting parameters to be in agreement with Higgs data and other
observables.

The paper is organized as follows. In Section~\ref{section0}, we will briefly review the $\mn$ and its relevant parameters for our analysis of the neutrino/sneutrino sector,
emphasizing the special role of the sneutrino in this scenario since its couplings have to be chosen so that the neutrino oscillation data are reproduced. 
In Section~\ref{section2},
we will introduce the phenomenology of the
$\widetilde{\nu}_{\tau}$ LSP, studying its pair production channels at the LHC, as well as the signals. These consist of two dileptons or a dilepton plus MET from the sneutrino decays.
Then, we will consider the existing dilepton displaced-vertex searches, and discuss its feasibility and significance on $\widetilde{\nu}_{\tau}$ 
searches.
In Section~\ref{methodology}, we will discuss the strategy that we employed to
perform scans searching for points of the parameter space of our scenario compatible with current experimental data on neutrino and Higgs physics, as well as flavor observables.
The results of these scans will be presented 
in Section~\ref{results-scans}, and applied to show the 
current reach of the LHC search on the parameter space of the $\widetilde{\nu}_{\tau}$ LSP based on the ATLAS 
8-TeV~result~\cite{Aad:2015rba}, and the prospects for the 13-TeV searches.
Finally, 
our conclusions are left for Section~\ref{Conclusions}.

\section{The $\mn$}
\label{section0}

The $\mn$~\cite{LopezFogliani:2005yw,Escudero:2008jg} is 
a natural extension of the MSSM where the $\mu$ problem is solved and, simultaneously, the neutrino data can be 
reproduced~\cite{LopezFogliani:2005yw,Escudero:2008jg,Ghosh:2008yh,Bartl:2009an,Fidalgo:2009dm,Ghosh:2010zi}. This is obtained through the presence of trilinear 
terms in the superpotential involving right-handed neutrino superfields $\hat\nu^c_i$, which relate the origin of the $\mu$-term to the origin of neutrino masses and mixing. 
The simplest superpotential of the $\mn$~\cite{LopezFogliani:2005yw,Escudero:2008jg,Ghosh:2017yeh} with three right-handed neutrinos is the following: 
\bea
W &=&
\epsilon_{ab} \left(
Y_{e_{ij}}
%Y^e_{ij} 
\, \hat H_d^a\, \hat L^b_i \, \hat e_j^c +
Y_{d_{ij}} 
%Y^d_{ij} 
\, 
%\delta_{\alpha\beta}\, 
\hat H_d^a\, \hat Q^{b}_{i} \, \hat d_{j}^{c} 
+
Y_{u_{ij}} 
%Y^u_{ij} 
\, 
%\delta_{\alpha\beta}\, 
\hat H_u^b\, \hat Q^{a}
%_{i\alpha} 
\, \hat u_{j}^{c}
\right)
\nonumber\\
% &+&
% \epsilon_{ab} Y^{\nu}_{ij} \, \hat H_u^b\, \hat L^a_i \, \hat \nu^c_j -
&+& 
\epsilon_{ab} \left(
Y_{{\nu}_{ij}} 
%Y^{\nu}_{i} 
\, \hat H_u^b\, \hat L^a_i \, \hat \nu^c_j
-
%\epsilon_{ab}
\lambda_i \, \hat \nu^c_i\, \hat H_u^b \hat H_d^a
\right)
+
\frac{1}{3}
\kappa_{ijk}
\hat \nu^c_i\hat \nu^c_j\hat \nu^c_k\,,
\label{superpotential}
\eea
where the summation convention is implied on repeated indices, with 
$a,b=1,2$ $SU(2)_L$ indices
and $i,j,k=1,2,3$ the usual family indices of the standard model (SM).

The simultaneous presence of the last three terms in 
Eq.~\eqref{superpotential} makes it impossible to assign $R$-parity charges consistently to the 
right-handed neutrinos ($\nu_{iR}$), thus producing explicit RPV (harmless for proton decay). Note nevertheless, that in the limit
$Y_{{\nu}_{ij}} 
\to 0$, $\hat \nu^c$ can be identified in the superpotential 
%of Eq.~(\ref{superpotential}) 
as a
pure singlet superfield without lepton number, similar to the 
next-to-MSSM (NMSSM)~\cite{Ellwanger:2009dp}, and therefore $R$ parity is restored.
%NMSSM
%where one extra singlet is added to the spectrum of the MSSM and $R_p$ is not broken.
Thus, the neutrino Yukawa couplings $Y_{\nu_{ij}}$ are the parameters which control the amount of RPV in the $\mn$, and as a consequence
%in the superpotential of Eq.~(\ref{superpotential}), 
this violation is small.
 After the electroweak symmetry breaking (EWSB)
induced by 
the soft SUSY-breaking terms of the order of the TeV, and 
 with the choice of CP conservation, 
the neutral Higgses ($H_{u,d}$) and right ($\widetilde \nu_{iR}$) and 
left ($\widetilde \nu_i$) sneutrinos 
develop the following vacuum expectation values (VEVs): 
\begin{eqnarray}
\langle H_{d}\rangle = \frac{v_{d}}{\sqrt 2},\quad 
\langle H_{u}\rangle = \frac{v_{u}}{\sqrt 2},\quad 
\langle \widetilde \nu_{iR}\rangle = \frac{v_{iR}}{\sqrt 2},\quad 
\langle \widetilde \nu_{i}\rangle = \frac{v_{i}}{\sqrt 2},
\end{eqnarray}
where $v_{iR}\sim$ TeV, {whereas 
%$v_i\sim Y_{\nu} v_u\lsim 10^{-4}$ GeV
$v_i\sim 10^{-4}$ GeV} because of the small contributions 
$Y_{\nu} \lsim 10^{-6}$
%to the left-sneutrino minimization equations, 
whose size is determined by the electroweak-scale 
seesaw of the $\mn$~\cite{LopezFogliani:2005yw, Escudero:2008jg}.
Note in this sense that the last term in 
Eq.~\eqref{superpotential} generates dynamically Majorana masses,
$m_{{\mathcal M}_{ij}}={2}\kappa_{ijk} \frac{v_{kR}}{\sqrt 2}\sim$ TeV.
On the other hand, the fifth term in the superpotential generates the $\mu$-term,
%the $\mu$-term,
$\mu=\la_i \frac{v_{iR}}{\sqrt 2}\sim$ TeV.

The new couplings and sneutrino VEVs in the $\mn$ induce new mixing of states.
%, and in particular there are eight neutral scalars and seven neutral pseudoscalars (Higgses-sneutrinos).
The associated mass matrices were studied in detail in
Refs.~\cite{Escudero:2008jg,Bartl:2009an,Ghosh:2017yeh}.
Summarizing, 
%in the case of one $\hat\nu^c$
%right-handed neutrino superfield, 
there are 
eight neutral scalars and seven neutral pseudoscalars (Higgses-sneutrinos),
eight charged scalars (charged Higgses-sleptons),
five charged fermions (charged leptons-charginos), and
ten neutral fermions (neutrinos-neutralinos). 
%{In our analysis of the electroweak sector below, we are mainly interested in the scalars/pseudoscalars and neutral fermions.} 
In the following, we will concentrate in briefly reviewing the neutrino and left sneutrino mass eigenstates, which are the relevant ones for our analysis.

The neutral fermions have the flavor 
composition 
$(\nu_{i},\widetilde B,\widetilde W,\widetilde H_{d},\widetilde H_{u},\nu_{iR})$. Thus,
with the low-energy bino and wino soft masses, $M_1$ and $M_2$, of the order of the TeV, and similar values for $\mu$ and $m_\mathcal{M}$ as discussed above, this generalized seesaw
%mixing left and right-handed neutrinos with neutralinos
produces three light neutral fermions dominated by the left-handed neutrino ($\nu_i$) flavor composition. 
In fact,
%Because of this structure, 
data on neutrino physics~\cite{Capozzi:2017ipn,deSalas:2017kay,deSalas:2018bym,Esteban:2018azc} can easily be reproduced at tree level~\cite{LopezFogliani:2005yw,Escudero:2008jg,Ghosh:2008yh,Bartl:2009an,Fidalgo:2009dm,Ghosh:2010zi}, even with diagonal Yukawa couplings~\cite{Ghosh:2008yh,Fidalgo:2009dm}, i.e.
$Y_{{\nu}_{ii}}=Y_{{\nu}_{i}}$ and vanishing otherwise.
A simplified formula 
for the effective mixing mass matrix of the 
light neutrinos is~\cite{Fidalgo:2009dm}:
\begin{eqnarray}
\label{Limit no mixing Higgsinos gauginos}
(m_{\nu})_{ij} 
\simeq \frac{Y_{{\nu}_{i}} Y_{{\nu}_{j}} v_u^2}
{6\sqrt 2 \kappa v_{R}}
                   (1-3 \delta_{ij})-\frac{v_{i} v_{j}}{4M^{\text{eff}}}
%&  
-\frac{1}{4M^{\text{eff}}}\left[\frac{v_d\left(Y_{{\nu}_{i}}v_{j}
   +Y_{{\nu}_{j}} v_{i}\right)}{3\lambda}
   +\frac{Y_{{\nu}_{i}}Y_{{\nu}_{j}} v_d^2}{9\lambda^2 }\right],
%   \nonumber\\
  \end{eqnarray}     
with
\begin{eqnarray}
\label{effectivegauginomass}
 M^{\text{eff}}\equiv M -\frac{v^2}{2\sqrt 2 \left(\kappa v_R^2+\lambda v_u v_d\right)
        \ 3 \lambda v_R}\left(2 \kappa v_R^{2} \frac{v_u v_d}{v^2}
        +\frac{\lambda v^2}{2}\right),
%\nonumber\\
\end{eqnarray} 
and
\begin{eqnarray}
\label{effectivegauginomass2}
\frac{1}{M} = \frac{g'^2}{M_1} + \frac{g^2}{M_2},
%\nonumber\\
\end{eqnarray} 
where
%$M_{1,2}$ are the bino and wino soft masses, and
$v^2 = v_d^2 + v_u^2 + \sum_i v^2_{i}={4 m_Z^2}/{(g^2 + g'^2)}\approx$ (246 GeV)$^2$.
%${M}= \frac{M_1 M_2}{g'^2 M_2 + g^2 M_1}$. 
For simplicity, we are also assuming in these formulas, and in what follows, $\lambda_i = \lambda$, $v_{iR}= v_{R}$, and
$\kappa_{iii}\equiv\kappa_{i}=\kappa$ and vanishing otherwise.
We
are then left with  
the following set of variables as independent parameters in the neutrino sector:
\bea
\lambda, \, \kappa,\, Y_{\nu_i}, \tan\beta, \, v_{i}, \, v_R, \, M,
\label{freeparameters}
\eea
where $\tan\beta \equiv v_u/v_d$
%$\tan\beta\equiv\frac{v_u}{v_d}$ 
and since $v_{i} \ll v_d, v_u$, we have
$v_d\approx v/\sqrt{\tan^2\beta+1}$.
{For the discussion, hereafter we will use indistinctly the subindices (1,2,3) $\equiv$ ($e,\mu,\tau$). {In the numerical analyses of the next sections, it will be enough for our purposes to consider
%we will adopt, without
%the loss of generality, 
the sign convention where all these parameters are positive.}
Of the five terms in Eq.~(\ref{Limit no mixing Higgsinos gauginos}),
%$(m^{\text{eff}}_{\nu})_{ij}$, 
the first two 
are generated through the mixing 
%of the left-handed neutrinos 
of $\nu_i$ with 
%the right-handed neutrinos 
$\nu_{iR}$-Higgsinos, and the rest of them
%The rest of them 
also include the mixing with the gauginos.
{These are the so-called $\nu_{R}$-Higgsino seesaw and gaugino seesaw, respectively~\cite{Fidalgo:2009dm}.}
}

As we can understand from these equations, neutrino physics in the $\mn$ is
closely related to the parameters and VEVs of the model, 
since the values chosen for them must reproduce current data on neutrino masses
and mixing angles.

%{One neutrino gets its mass at tree level, whereas the other two at one loop. As discussed in Ref~\cite{Lara:2018rwv}, the tree-level mass can be approximated
%as $m_{\nu} \approx {\sum_i {v_{i}^2}}/{4M}$, with
%$\frac{1}{M}\equiv\frac{g'^2}{M_1} + \frac{g^2}{M_2}$.
%}
%and in agreement with experimental constraints on neutrino masses and mixing angles; 
%The rest of neutral fermions get masses around the TeV scale.
%However, if $M_1$ is small compared with the rest of the parameters, the fourth lightest eigenstate of the mass matrix, which we identify as the lightest neutralino {$\tilde{\chi}^0_1$}, is mainly bino dominated and the LSP 
%Note in this sense that Higgsino and right-handed neutrino entries in this matrix are proportional to the right sneutrino VEVs, that are expected to be large from the minimization conditions when the soft scalar masses and trilinear parameters are in the range of the TeV. 
%with {$m_{\tilde{\chi}^0_1}\approx M_1$}, since the largest off-diagonal mass entry $m_{\tiny{\tilde B \tilde H_u}}=\frac{1}{\sqrt{2}}g'v_u$ is small. 

Concerning the neutral scalars in the $\mn$, although they have 
flavor composition 
($H_d^{\mathcal{R}}, H_u^{\mathcal{R}}, 
\widetilde\nu^{\mathcal{R}}_{iR},\widetilde\nu^{\mathcal{R}}_{i}$), 
the off-diagonal terms of the mass matrix mixing the left sneutrinos with Higgses and right sneutrinos are suppressed by $Y_{\nu}$ and $v_{iL}$, implying that the left sneutrino states will be almost pure.
The same happens for the pseudoscalar left sneutrino states $\widetilde{\nu}^{\mathcal{I}}_{i}$, which have in addition degenerate masses with the scalars 
{$m_{\widetilde{\nu}^{\mathcal{R}}_{i}}
\approx
 m_{\widetilde{\nu}^{\mathcal{I}}_{i}}
\equiv 
m_{\widetilde{\nu}_{i}}$. 
{From the minimization equations for $v_i$, we can write their approximate tree-level values as
\bea
m_{\widetilde{\nu}_{i}}^2
\approx  
\frac{Y_{{\nu}_i}v_u}{v_i} \frac{v_R}{\sqrt 2}
\left[
\frac{-T_{{\nu}_i}}{Y_{{\nu}_i}}
+ \frac{v_R}{\sqrt 2} \left(-\kappa
+
\frac{3\lambda}{\tan\beta}\right)
\right],
\label{evenLLL2}
\eea
where $T_{{\nu}_i}$ are the trilinear parameters in the soft Lagrangian, $-\epsilon_{ab} T_{{\nu}_{ij}} H_u^b \widetilde L^a_{iL} \widetilde \nu_{jR}^*$, taking for simplicity $T_{{\nu}_{ii}}=T_{{\nu}_i}$ and vanishing otherwise. Therefore, left sneutrino masses introduce in addition to the parameters of Eq.~(\ref{freeparameters}), the 
\bea
T_{{\nu}_i},
\label{tia}
\eea
as other relevant parameters for our analysis.}
{In the numerical analyses of Sections~\ref{methodology} and~\ref{results-scans}, we will use negative values for them in order to avoid tachyonic left sneutrinos.}

Since we have assumed diagonal sfermion mass matrices, and
from the minimization conditions 
%of Eqs.~(\ref{tadpoles1})--(\ref{tadpoles4}), 
we have eliminated
the 
%low-energy 
soft masses $m_{H_{d}}^{2}$, $m_{H_{u}}^{2}$, $m^2_{\widetilde{\nu}_{iR}}$ and
$m^2_{\widetilde{L}_{iL}}$ in favor
of the VEVs, the parameters in Eqs.~(\ref{freeparameters}) and (\ref{tia}),
together with the rest of soft trilinear parameters, soft scalar masses, and soft gluino masses
\bea
%T_{{\nu}_i}, \, 
T_{\lambda}, \, T_{\kappa}, \, T_{u_{i}}, \, T_{d_{i}}, \, T_{e_{i}}.
 \, m_{\tilde Q_{iL}},\, 
m_{\tilde u_{iR}}, \, m_{\tilde d_{iR}}, \,
m_{\tilde e_{iR}}, \,
%A^u_{ij}, \, A^d_{ij}, \, A^e_{ij}, \, 
M_3,
\label{freeparameterssoft}
\eea
constitute our whole set of free parameters, and are specified at low scale.
Note that the parameters $\ka$, $v_R$ and $T_{\kappa}$ are the key ingredients to determine
the mass scale of the right sneutrino 
states \cite{Escudero:2008jg,Ghosh:2008yh}.
For example, for $\lambda\lsim 0.01$ they are free from any doublet contamination, and 
the masses can be approximated by~\cite{Ghosh:2014ida,Ghosh:2017yeh}:
\bea
m^2_{\widetilde{\nu}^{\mathcal{R}}_{iR}} \approx   \frac{v_R}{\sqrt 2}
\left(T_{\kappa} + \frac{v_R}{\sqrt 2}\ 4\kappa^2 \right), \quad
m^2_{\widetilde{\nu}^{\mathcal{I}}_{iR}}\approx  - \frac{v_R}{\sqrt 2}\ 3T_{\kappa}.
\label{sps-approx2}
\eea 
Thus {we will use negative values for $T_{\kappa}$ in order to avoid tachyonic pseudoscalar right sneutrinos.}
{Given that we will focus on a $\widetilde{\nu}_{\tau}$ LSP with a mass smaller than 100 GeV, we will also use negative values for $T_{u_3}$ in order to avoid too light 
%tachyonic 
left sneutrinos due to loop corrections.}

{Let us finally point out, that if we follow the usual assumption based on the breaking of supergravity, that all the trilinear parameters are proportional to their corresponding Yukawa couplings, defining $T_{\nu}= A_{\nu} Y_{\nu}$ we can write Eq.~(\ref{evenLLL2}) as: 
\bea
m_{\widetilde{\nu}_{i}}^2
\approx  
\frac{Y_{{\nu}_i}v_u}{v_i} \frac{v_R}{\sqrt 2}
\left[
-A_{{\nu}_i}
+ \frac{v_R}{\sqrt 2} \left(-\kappa
+
\frac{3\lambda}{\tan\beta}\right)
\right],
\label{evenLLL22}
\eea
and the parameters $A_{{\nu}_i}$ substitute the $T_{{\nu}_i}$ as the most representative.
We will use both type of parameters throughout this work.}

%%%%%%%%%%%%%%%%%%%%%%%%%%%%%%%%%%%%%%%%%
\subsection{Neutrino/sneutrino physics}
\label{neusneu}
%%%%%%%%%%%%%%%%%%%%%%%%%%%%%%%%%%%%%%%%

{Since reproducing neutrino data is an important asset of the 
$\mn$, as explained above, we will try to establish here qualitatively what regions of the parameter space are
the best in order to be able to obtain correct neutrino masses and mixing angles. In particular, we will determine natural hierarchies among 
neutrino Yukawas, and among left sneutrino VEVs.}

In addition, left sneutrinos are special in the $\mn$ with respect to other SUSY models. This is because, as 
discussed in Eq.~(\ref{evenLLL2}), their masses are determined by the minimization equations with respect to
$v_i$. {Thus, they depend not only on left sneutrino VEVs but also on neutrino Yukawas, and as a consequence neutrino physics is very relevant.
In particular, if we work with Eq.~(\ref{evenLLL22}) assuming the simplest situation that all the $A_{{\nu}_i}$ are naturally of the order of the TeV, neutrino physics determines sneutrino masses through the prefactor
${Y_{{\nu}_i}v_u}/{v_i}$.
%this hierarchy of Yukawas and VEVs determines from
%Eq.~(\ref{evenLLL22}) that 
%$m_{\widetilde{\nu}_{1}}$ is the smallest of all sneutrino masses.
Considering the normal ordering (NO) for the neutrino mass spectrum, which is nowadays favored by the analyses of neutrino data \cite{Capozzi:2017ipn,deSalas:2017kay,deSalas:2018bym,Esteban:2018azc}, representative solutions for neutrino/sneutrino physics using diagonal neutrino Yukawas in this scenario are summarized below. Note that these solutions take advantage of the 
dominance of the gaugino seesaw for some of the three neutrino families.}

1) $M<0$, with $Y_{\nu_1}<Y_{\nu_2}, Y_{\nu_3}$, and $v_1 > v_2, v_3$.

\noindent 
As explained in Refs.~\cite{Fidalgo:2009dm,Gomez-Vargas:2016ocf}, a negative value for $M$ is useful in order to reproduce neutrino data with $Y_{\nu_1}$ the smallest Yukawa and $v_1$ the largest VEV. 
{Essentially, this is because 
a small tuning in 
Eq.~(\ref{Limit no mixing Higgsinos gauginos}) 
between the gaugino seesaw and the $\nu_{R}$-Higgsino seesaw is
necessary in order to obtain the correct mass of the first family.
Here the contribution of the gaugino seesaw is always the largest one. 
On the contrary, for the other two neutrino families, the contribution of the $\nu_{R}$-Higgsino seesaw is the most important one and that of the gaugino seesaw is less relevant for the tuning.}
{Following the above discussion about the prefactor of Eq.~(\ref{evenLLL22}), these hierarchies of Yukawas and VEVs determine 
%from Eq.~(\ref{evenLLL22}) 
that 
$m_{\widetilde{\nu}_{1}}$ is the smallest of all the sneutrino masses.
}

{
2) $M>0$, with $Y_{\nu_3} < Y_{\nu_1} < Y_{\nu_2}$, and $v_1<v_2\sim v_3$.}

\noindent 
{In this case, it is easy to find solutions with the gaugino seesaw as the dominant one for the third family. Then, $v_3$ determines the corresponding neutrino mass and $Y_{\nu_3}$ can be small.
On the other hand, the NO for neutrinos determines that the first family dominates the lightest mass eigenstate implying that $Y_{\nu_{1}}< Y_{\nu_{2}}$ and $v_1 < v_2,v_3$, {with both $\nu_{R}$-Higgsino and gaugino seesaws contributing significantly to the masses of the first and second family}. Taking also into account that the composition of these two families in the second mass eigenstate is similar, we expect $v_2 \sim v_3$. 
%\R{The value of $Y_{\nu_2}$ is important in order to have a complete agreement in admixtures and mass differences with the experiments.}
Now for this solution we will have $m_{\widetilde{\nu}_{3}}$ as the smallest of all the sneutrino masses.}

%El más ligero es la  tercera familia de los sneutrinos si domina el gaugino see-saw.  Cuando forzamos que sea la tercera familia sea la LSP, $Y_nu3$  pequeño esta  favoreciendo. 
%In this case, $m_{\widetilde{\nu}_{3}}$ as the smallest of all sneutri

3) $M>0$, with $Y_{\nu_2} < Y_{\nu_1} < Y_{\nu_3}$, and $v_1<v_2\sim v_3$.

\noindent 
These solutions can be deduced from the previous ones {in 2)} interchanging the values of the third family, $Y_{\nu_3}$ and $v_3$, with the corresponding ones of the second family, $Y_{\nu_2}$ 
and $v_2$.
A small adjust in the parameters will lead again to a point in the parameter space satisfying neutrino data. This is clear from the fact that $\theta_{13}$ and $\theta_{12}$ are not going to be significantly altered, whilst $\theta_{23}$ may require a small tuning in the parameters. 
If the gaugino seesaw dominates for the second family, $v_2$ determines the corresponding neutrino mass and $Y_{\nu_2}$ can be small. Then,
$m_{\widetilde{\nu}_{2}}$ will be the smallest of all sneutrino masses.

%\bl{4) Dominance of the $\nu_\text{R}$-Higgsino see-saw.}

%\noindent 
%\bl{
%We already mentioned that the first family dominates the lightest mass eigenstate in the normal hierarchy, implying in this case that $Y_{\nu_1}$ is the smallest Yukawa. 
%\R{Daniel, aqui no debería ir una discusion como se hace en los puntos anteriores?.}
%}

\vspace{0.2cm}

{We will see in the next subsection that solutions of type 2) are the ones interesting for our analysis.
%natural and simplest one when sneutrinos left of the third familly are the lightest of the sneutrinos left.
}

{
Let us finally point out that when off-diagonal neutrino Yukawas are allowed, it is not possible to arrive to a general conclusion regarding the hierarchy in sneutrinos masses, specially when the gaugino seesaw is sub-dominant. This is because one can play with the hierarchies among $v_i$ with enough freedom in the neutrino Yukawas in order to reproduce the 
experimental results. Therefore, there is no a priori knowledge of the hierarchies in 
the sneutrino masses, and carrying out an analysis case by case turns out to be
necessary. 
}
%%%%%%%%%%%%%%%%%%%%%%%%%%%%%%%%%%%%%%%%%
\subsection{$\widetilde{\nu}_{\tau}$ LSP}
%%%%%%%%%%%%%%%%%%%%%%%%%%%%%%%%%%%%%%%%%

In the $\mn$, because of RPV any SUSY particle can be a candidate for the LSP.
Nevertheless, the case of the $\widetilde{\nu}_{\tau}$
LSP turns out to be particularly interesting because of the large value of the tau Yukawa coupling, 
which can give rise to significant BRs for 
decays to\footnote{In what follows, the symbol $\ell$ will be used for an electron or a muon,
{$\ell=e,\mu$,}
and charge conjugation of fermions is to be understood where appropriate.}
$\tau\tau$ and $\tau \ell$, once the sneutrinos are dominantly pair-produced via a Drell-Yan process mediated by a virtual
$W$, $Z$ or $\gamma$, as we will discuss in the next section.

{There is enough freedom in the parameter space of the $\mn$ 
%trilinear 
%soft 
%parameters 
%$A_{\nu}$ 
in order to get light left sneutrinos.
Assuming as discussed above that the $A_{{\nu}_i}$ are 
%in Eq.~(\ref{evenLLL22}) 
naturally of the order of the TeV, values of the prefactor of
Eq.~(\ref{evenLLL22})
${Y_{{\nu}_i}v_u}/{v_i}$ in the range of about $0.01-1$, i.e.
$Y_{{\nu}_i}\sim 10^{-8}-10^{-6}$, will give rise to left sneutrino masses in the
range of about $100-1000$ GeV.
Thus, with the hierarchy of neutrino Yukawas 
$Y_{{\nu}_{3}}\sim 10^{-8}-10^{-7}<Y_{{\nu}_{1,2}}\sim 10^{-6}$, we can obtain a
 $\widetilde{\nu}_{\tau}$ LSP with a mass around 100 GeV whereas the masses of
 $\widetilde{\nu}_{e,\mu}$ are of the order of the TeV.
Clearly, we are in the case of solutions for neutrino physics of type 2) discussed in Subsection~\ref{neusneu}. 
{Actually this type of hierarchy, with significant values for $Y_{{\nu}_{1,2}}$,
increases the dilepton BRs of the $\widetilde{\nu}_{\tau}$ LSP producing signals that can be probed at the LHC, as the analysis of the next sections 
%Section~\ref{results-scans} 
will show.
%where many solutions of this kind are found in the parameter space of the $\mn$. 
}

{It is worth noticing here that in this scenario the left stau can be naturally the NLSP, since it is only a little heavier than the $\widetilde{\nu}_{\tau}$ because they are in the same $SU(2)$ doublet, with the mass splitting mainly 
due to the usual small D-term contribution, $-m_W^2 \cos 2\beta$.
As we will see in the next section, this has implications for the production of the left sneutrino LSP at 
the LHC, because the direct production of sleptons and their decays is a significant source of sneutrinos.
}

\section{Searching for $\widetilde{\nu}_{\tau}$ LSP at the LHC
%Tau left sneutrino LSP phenomenology
}
\label{section2}

To probe the $\widetilde{\nu}_{\tau}$
%tau left sneutrino 
LSP, the dilepton displaced-vertex
searches are found to be the most promising. 
Following the strategy of Ref.~\cite{Lara:2018rwv},
we will compare the 
%$\mn$ 
predictions of our current scenario with three right-handed neutrinos
%for a  
%$\widetilde{\nu}_{\tau}$ LSP 
with the ATLAS search~\cite{Aad:2015rba} for long-lived
particles using displaced ($\gsim$ 1 mm) lepton pairs $\ell \ell$ in $pp$ collisions at
$\sqrt s= 8$~TeV, as well as the prospects for the 13-TeV searches.
%, which allows us to constrain the parameter space of the model.

%%%%%%%%%%%%%%%%%%%%%%%%%%%%%%%%%%%%%%%%%%%%%%%%%%%%%%%%%%%%%%%%
\begin{figure}[t!]
\centering
\subcaptionbox{ 
\label{fig:production31a} $Z$ channel}{\includegraphics[scale=0.45]{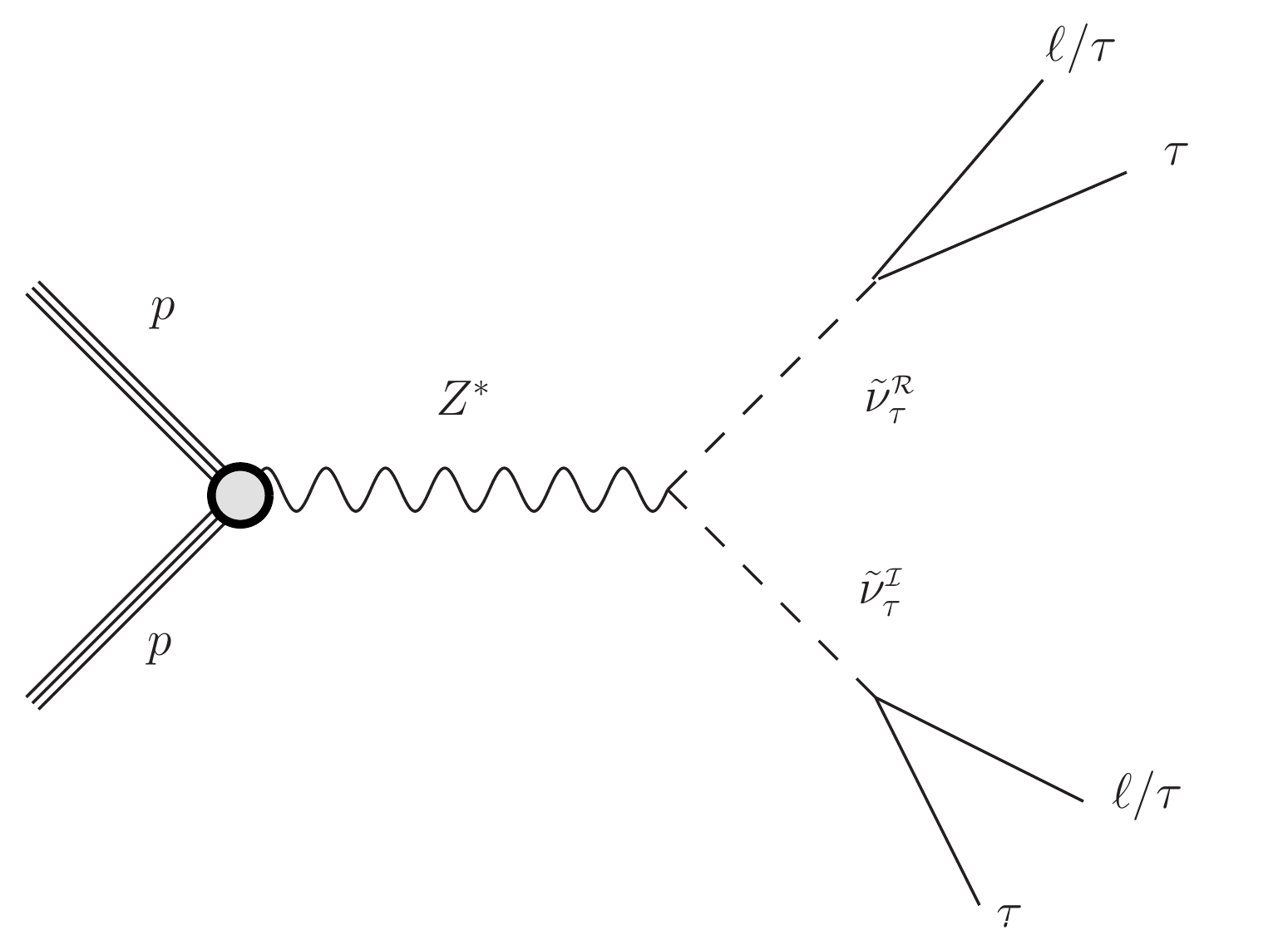}} 
  \subcaptionbox{
\label{fig:production31c} $\gamma ,Z$ channels}{\includegraphics[scale=0.45]{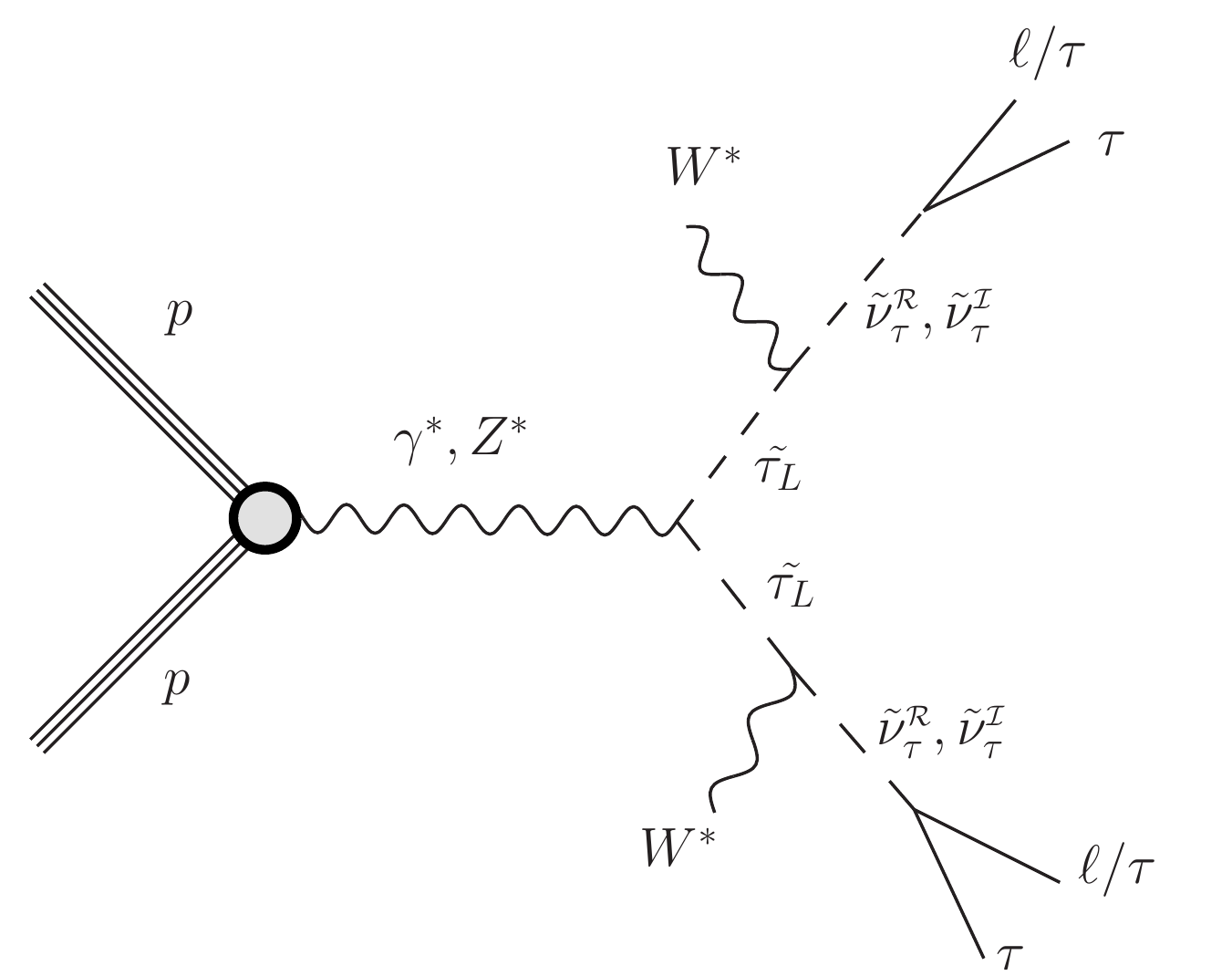}}
  \subcaptionbox{
\label{fig:production31b} $W$ channel}{\includegraphics[scale=0.45]{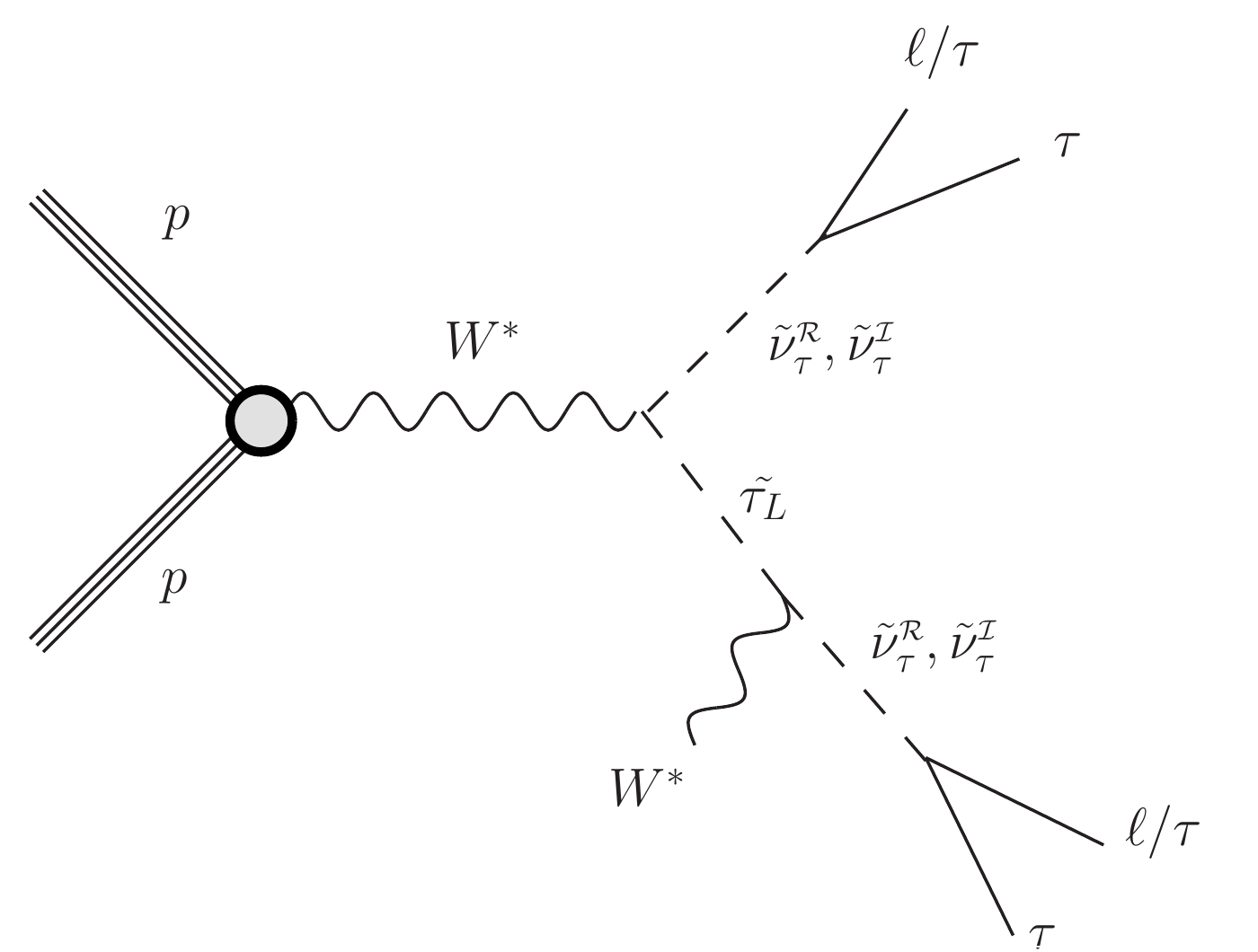}}
\caption{Decay channels into two $\tau\,\ell/\tau$, from a pair production at the LHC of scalar and pseudoscalar tau left sneutrinos co-LSPs.
{Decay channels into one $\tau\,\ell/\tau$ plus neutrinos are the same but substituting 
in (a), (b) and (c) one of the two vertices by a two-neutrino vertex}.
}
\label{fig:Tau-left-sneutrino-production.png}
\end{figure} 
%%%%%%%%%%%%%%%%%%%%%%%%%%%%%%%%%%%%%%%%%%%%%%%%%%%%%%%%%%%%%%%%%

The direct production of $\widetilde{\nu}_{\tau}$
occurs via a $Z$ channel giving rise to a pair of scalar and pseudoscalar left sneutrinos, as shown in Fig.~\ref{fig:Tau-left-sneutrino-production.png}(a). 
Note that they are co-LSPs since they have essentially degenerate masses, as explained in the previous section.
On the other hand, since the left stau is typically the NLSP its direct production and decay is another important source of the $\widetilde{\nu}_{\tau}$ LSP.
In particular, pair production can be obtained
through a $\gamma$ or $Z$ decaying into two staus,
as shown in Fig.~\ref{fig:Tau-left-sneutrino-production.png}(b),
with the latter having a dominant RPC prompt decay into 
a (scalar or pseudoscalar) sneutrino plus an off-shell $W$ producing a
soft meson or a pair of a charged lepton and a neutrino.
%, which are usually undetectable. 
Note that although RPV decays of the stau are possible, e.g. stau into a tau plus a neutrino, they are extremely supressed compared to the RPC one. Numerically, the stau has partial decay widths through RPV diagrams $\sim 10^{-14} - 10^{-13}$ GeV, while the ones corresponding to the RPC three-body decays are $\sim 10^{-7}$ GeV. Therefore, its proper decay length is $\sim 10^{-9}$ m, with the BRs corresponding to the RPV decays $< 10^{-6}$.
Sneutrinos can also be pair produced through a 
$W$ decaying into a stau and a (scalar or pseudoscalar) sneutrino
as shown in Fig.~\ref{fig:Tau-left-sneutrino-production.png}(c), with the stau 
decaying as before.
% Notice that in the last two processes a pair of scalar (or pseudoscalar) sneutrinos can be produced unlike the process in Fig.~\ref{fig:production31a}, where always a pair of scalar and pseudoscalar sneutrinos is produced.

Subsequently, the pair-produced $\widetilde{\nu}_{\tau}$ can decay
into $\tau\,\ell/\tau$.
As a result of the mixing between left sneutrinos and Higgses, 
the sizable decay of $\widetilde{\nu}_{\tau}$ into $\tau\tau$ is possible
because of the large value of the tau Yukawa coupling.  
Other sizable decays into $\tau\,\ell/\tau$ can occur through the Yukawa interaction of 
$\widetilde{\nu}_{\tau}$ with $\tau$ and charged Higgsinos,
via the mixing between the latter and $\ell$ or $\tau$.
 %with the latter decaying  as well as the decay into  
%and leptons and charged Higgsinos,
%a detectable decay of the $\widetilde{\nu}_{\tau}$ 
%into $\tau\,\ell/\tau$ is possible.
%In Fig.~\ref{fig:production31}, we show the decays into a pair of $\tau\,\tau/\ell$.
To analyze these processes we can write approximate formulas for
the partial decay widths of the scalar/pseudoscalar tau left sneutrino.
The one
%The partial decay width 
%of the scalar/pseudoscalar tau left sneutrino 
%$\widetilde{\nu}^\mathcal{R/I}_{\tau L}$ 
into 
$\tau\tau$ is given by: %corresponding to the diagrams of 
% Figs.~\ref{fig:efflambda-to-leptons-HuHd2} and~\ref{fig:eff-to-leptons-higgsino2}, 
%can be approximated as (\R{CHECK}):
%
\begin{equation}
  \Gamma \left(\widetilde{\nu}_{\tau}
\rightarrow 
 \tau\tau\right) 
\approx\frac{m_{\widetilde{\nu}_{\tau}
}}{{16\pi}}
 \left(Y_\tau 
 Z^{H/A}_{\widetilde{\nu}_{\tau} H_d}
% Z^{H(A)}_{\widetilde \nu^\Re_\tau ( \widetilde \nu^\Im_\tau) H_d}  
%- Y_{\nu_{\tau,j}} \frac{Y_\tau v_{Rj}}{\sum_{l=1}^3{\lambda_l v_{Rl}}}\right)^2,
%\left(Y_{\tau} Z^{H/A}_
%{\widetilde{\nu}_{\tau} H_d}
- Y_{\nu_{\tau}} 
\frac{Y_{\tau}
}{3\lambda}
\right)^2,
  \label{eq:3.220}
\end{equation}
where $Y_{\tau}\equiv Y_{e_{33}}$, 
%$Y_{\nu_{\tau}}\equiv Y^{\nu}_{3}$, 
and $Z^{H/A}$ is the matrix which diagonalizes the mass matrix for the neutral 
scalars/pseudoscalars.  
%($H_d, H_u, \widetilde\nu_{R},
%\widetilde\nu_{i}$)~\cite{Escudero:2008jg}. 
%~\cite{Escudero:2008jg,Ghosh:2017yeh}. 
The latter is
determined by the neutrino Yukawas, which are the order parameters of
the RPV. The contribution of $\lambda$ in the second term of
Eq.~(\ref{eq:3.220}) is due to the charged Higgsino mass that can be
approximated by the value of 
$\mu=3\lambda \frac{v_{R}}{\sqrt 2}$. The partial decay width into $\tau
\ell$ can then be approximated for both sneutrino states by the second
term of Eq.~(\ref{eq:3.220}) with the substitution
$Y_{\nu_{\tau}}\rightarrow Y_{\nu_{\ell}}$: 
\begin{equation}
%  \Gamma (\widetilde{\nu}^\mathcal{R/I}_{3 L}\rightarrow 
%  e^{+}_{3L} e^{-}_{jL}\ , e^{-}_{3R} e^{+}_{jR}) 
  \Gamma \left(\widetilde{\nu}_{\tau}
%^\mathcal{R/I}_{\tau L}
\rightarrow 
\tau \ell\right) 
\approx\frac{m_{\widetilde{\nu}_{\tau}
%\widetilde{\nu}
}}{{16\pi}}
% \left(
 %Y_\tau Z^{H(A)}_{\widetilde \nu^\Re_\tau ( \widetilde \nu^\Im_\tau) H_d} 
 %-
 %Y_{\nu_{\tau,j}} \frac{Y_\tau v_{Rj}}{\sum_{l=1}^3{\lambda_l %v_{Rl}}}\right)^2.
\left(Y_{\nu_{\ell}}
\frac{Y_{\tau} 
}{3\lambda}
\right)^2.
  \label{eq:3.22}
\end{equation}

{On the other hand, 
the gauge interactions of 
$\widetilde{\nu}_{\tau}$ with neutrinos and binos (winos)
can produce a large decay width into neutrinos,
via the gauge mixing between these gauginos and neutrinos.}
% the mixing between neutrinos and Binos and Winos, 
% which is proportional to the gauge couplings,
% can produce as well a large decay width of 
% left sneutrinos (of any family)
% into neutrinos. 
This partial decay width
%corresponding to the diagrams of Fig.~\ref{fig:eff-to-neutrinos2}, 
can be approximated for scalar and pseudoscalar sneutrinos as
%\begin{equation}
%\sum_i \Gamma \left(\widetilde{\nu}_{\tau}
%\rightarrow  
% \nu_\tau \nu_i\right)
%\approx\frac{m_{\widetilde{\nu}_{\tau}}}{16\pi}
%\frac{1}{2M^2} \sum_i{v_{i}^2},
%  \label{eq:3.55}
%\end{equation}
\begin{eqnarray}
\sum_{i}{\Gamma(\widetilde \nu_\tau \to \nu_\tau \nu_i  )} \approx \frac{m_{\widetilde \nu_\tau}}{16\pi}
 \sum_{i}{\left|\frac{g'}{2}U^V_{i4}{-}\frac{g}{2}U^V_{i5}\right|^2},
% \nonumber
 \label{--sneutrino-decay-width-2nus}
\end{eqnarray}
where $U^V$ is the matrix which diagonalizes the mass matrix for the 
neutral fermions, and the above entries can 
%with $M$ a kind of average of Bino and Wino masses defined in 
%Eq.~(\ref{eq:3.550}).
be approximated as 
\begin{eqnarray}
U^V_{i4}\approx\frac{{-}g'}{\sqrt{2}M_1}\sum_{l}{v_{l}U^{PMNS}_{il}},\nonumber\\
U^V_{i5}\approx\frac{g}{\sqrt{2}M_2}\sum_{l}{v_{l}U^{PMNS}_{il}}.
%\nonumber
\label{--sneutrino-decay-width-2nus2}
\end{eqnarray}
Here $U^{PMNS}_{il}$ are the entries of the PMNS matrix, with
$i$ and $l$ neutrino physical and flavor indices, respectively. 
%runs over physical neutrinos and l sum over neutrino flavors. 
%, thus can be expressed in terms of neutrino mixing angles. For normal ordering:
The relevant diagrams for  
$\widetilde{\nu}_{\tau}$
%tau left sneutrino 
searches that include this decay mode are the same as in
Fig.~\ref{fig:Tau-left-sneutrino-production.png}, but 
%For decays into one $\tau\,\tau/\ell$ plus MET, the diagram is the same
substituting one of the $\tau\,\ell/\tau$
%$\tau\,\tau/\ell$ 
vertices by a two-neutrino vertex.

Let us remark that other decay channels of 
the $\widetilde{\nu}_{\tau}$ can be present and have been taken into account in our numerical computation, but they turn out to be negligible
%~\cite{Ghosh:2017yeh} 
for the sneutrino masses that we are interested in this work.

Given the above results valid for three families of right-handed neutrino superfields, we can now follow the prescription of Ref.~\cite{Lara:2018rwv} for improving and recasting the 
ATLAS search~\cite{Aad:2015rba} to the case of the $\widetilde{\nu}_{\tau}$.
One of the problems with the existing searches
\cite{Aad:2015rba, CMS:2014hka, Aaboud:2017iio, Aad:2019kiz} is that they are designed for a generic purpose and
therefore not optimized for light metastable particles such as 
the $\widetilde{\nu}_{\tau}$; we thus proposed in Ref.~\cite{Lara:2018rwv}
%tau left sneutrino. 
a strategy of improving these searches
by lowering trigger thresholds, relying on a high level trigger that
utilizes tracker information. 
This optimization turned out to be
quite feasible and considerably improves the sensitivity of the 
displaced-vertex searches to long-lived $\widetilde{\nu}_{\tau}$. 
In particular, in the ATLAS 8-TeV analysis, the events
must satisfy the following {trigger} requirements~\cite{Aad:2015rba}:
\begin{itemize}
 \item[$\bullet$] One muon with $p_{\rm T}>50$~GeV and
$|\eta|<1.07$, one electron
with $p_{\rm T}>120$~GeV or two electrons with $p_{\rm T}>40$~GeV each,
\end{itemize}
{and off-line selection requirements:}
\begin{itemize}
% \item[$\bullet$] Trigger muon must have  $|\eta|<1.07$.
 \item[$\bullet$] One pair $e^+e^-$, $\mu^+\mu^-$ or $e^\pm\mu^\mp$ 
with $p_{\rm T} >10$~GeV and $0.02<|\eta|<2.5$ for each particle.
\end{itemize}
As shown in Ref.~\cite{Lara:2018rwv}, 
the trigger requirement for electrons is so restrictive that it makes the selection efficiency for the dielectron channel be a few percent level, while for the $\mu^+ \mu^-$ and $e^\pm\mu^\mp$ channels the efficiency can be a few tens of percent. 
%, and thus ineffective for light sneutrino searches.
We can however overcome this difficulty by optimizing the trigger
requirements for left sneutrino searches by relaxing the momentum thresholds \cite{Lara:2018rwv}. 
%For instance, for the muon
%trigger, the ATLAS 8-TeV analysis uses only the muon spectrometer and
%requires $p_{\rm T} > 50$~GeV, as discussed above. On the other hand, 
In fact, it is possible to reduce momentum thresholds for triggers by means of established techniques. For instance, the {\tt mu24i} trigger used in the ATLAS experiment \cite{Aad:2014sca} only requires $p_{\rm T} >
24$~GeV; such a low threshold can be achieved thanks to the information from the inner detector. 
%, which is an isolated single muon trigger at
%the event-filter, also uses the information from the inner detector and
%requires the transverse momentum threshold of $p_{\rm T} >
%24$~GeV.
%\footnote{This 
%trigger should also satisfy a loose isolation selection, the sum of the
%$p_{\rm T}$ of tracks in a cone of $\Delta R<0.2$ centered around the
%muon candidate after eliminating the muon transverse momentum
%$(p_{\rm T})_\mu$ should be smaller than $0.12 \times (p_{\rm T})_\mu$;
%this requirement is so loose that almost all isolated muons from the
%$Z$-boson decays pass the criterion. Since the muons coming from the
%sneutrino decays are also expected to be isolated, we can expect that
%this requirement scarcely affects the sneutrino event selection. For
%this reason, we do not take account this effect in the following
%analysis.} 
This information can also improve the trigger performance 
%With the help of the inner detector information, this {\tt mu24i}
%trigger has a good performance 
in a wider range of the pseudorapidity of
tracks, and thus we can also relax the requirement on $\eta$; from
$|\eta| < 1.07$ to $|\eta| < 2.5$ \cite{Aad:2014sca}. It is then argued in Ref.~\cite{Lara:2018rwv} that we can still consider the number of background events to be zero even after we relax the momentum threshold. Consequently, 
to exploit this
trigger instead of that used in Ref.~\cite{Aad:2015rba} can
significantly enhance the sensitivity to light sneutrinos, since the
typical momentum of muons from the sneutrino decays is a few tens of
GeV. 
After all, one can use the following criteria for the optimized 8-TeV analysis:\footnote{We could have required a lower threshold for the electron trigger as well, but we do not consider this optimization since we are unable to estimate the increase in the number of background events caused by the relaxation in the trigger requirement \cite{Lara:2018rwv}. } 
\begin{itemize}
 \item[$\bullet$] At least one muon with $p_{\rm T}>24$~GeV.
 \item[$\bullet$] One pair $\mu^+\mu^-$ or $e^\pm\mu^\mp$ with $p_{\rm
	      T} >10$~GeV and $0.02<|\eta|<2.5$ for each particle.
\end{itemize}

We can also assume an optimization of the trigger requirements in the 13-TeV searches. It is again discussed in Ref.~\cite{Lara:2018rwv} that one
can use the following criteria for the 13-TeV analysis:
\begin{itemize}
 \item[$\bullet$] At least one electron or muon with $p_{\rm T}>26$~GeV.
 \item[$\bullet$] One pair $\mu^+\mu^-$, $e^+e^-$, or $e^\pm\mu^\mp$ with $p_{\rm
	      T} >10$~GeV and $0.02<|\eta|<2.5$ for each particle.
\end{itemize}
Since we do not have the 13-TeV result for dilepton displaced-vertex searches for the moment, we just assume the expected number of background events to be zero, which should be validated in the future experiments.
It is worth noticing that unlike the previous two trigger requirements,
in this case the $p_T$ threshold of 26 GeV is for both
muons and electrons. 
%This will be relevant in the analysis
%of Section~\ref{results-scans}, since the Yukawa of the electron neutrino %turns out to be significant and therefore its contribution to the
%value of the decay length can be important
%the contribution of the corresponding BR 
%(see Eq.~(\ref{eq:3.22})).
The improvement of the selection efficiencies, $\epsilon_{\rm sel}$,
for different masses, for the three production processes, 
and for the $\mu\mu$, $\mu e$, and $ee$ channels,
can be found in 
Tables III-IX of Ref.~\cite{Lara:2018rwv}.
As pointed out also in that work, this possible
improvement is not only for the ATLAS analysis but also for the CMS one
\cite{CMS:2014hka}.

%
%We will also consider this possibility in this work, as well as an optimization of the 13~TeV LHC searches, and will show the
%prospects for investigating the 
%$\mn$ 
%parameter space of our scenario by searching for the
%$\widetilde{\nu}_{\tau}$.
%tau left sneutrinos 
%at the 13~TeV LHC run. Possibilities of furtherimprovements for these searches will also be discussed.

We can now discuss how to obtain the limits for light sneutrinos. Throughout our analysis, we assume that the number of both background and signal events to be zero, as in the ATLAS 8-TeV search result \cite{Aad:2014sca}. 
The limits from the ATLAS search
can be translated into a vertex-level efficiency, taking into account
the lack of observation of events for any value of the decay
length. Therefore, $\epsilon_{\text{vert}}(c\tau)$ can be obtained as
the ratio of the number of signal events compatible with zero observed
events {(which in this case is 3 as we assume zero background)} and that corresponding to the upper limits given in
Ref.~\cite{Aad:2015rba} with an appropriate modification described in Ref.~\cite{Lara:2018rwv}; for example, we can use
{the purple-shaded solid line of} Fig.~3 in the later work
%Ref.~\cite{Lara:2018rwv}
to obtain the vertex-level efficiency
$\epsilon^{\mu\mu}_{\text{vert}}(c\tau)$ for the dimuon channel. It is found that the efficiency decreases significantly for $c \tau \lesssim 1$~mm, which has important implications for the prospects of the $\widetilde{\nu}_{\tau}$ searches as we will see below.  By multiplying the
number of the events passing the trigger and event selection criteria
%, which
%is discussed in Ref.~\cite{Lara:2018rwv}, 
with this vertex-level efficiency, we can estimate
the total number of signal events; for the 8-TeV 
case, {this is given 
for the $\mu\mu$ channel 
%,
%$e e$ and $e\mu$ channels
by
%\R{(check this new way of expressing it)}
% \begin{eqnarray}
% \mathcal{L}
% & \times &
% \left[\sigma(pp\to Z^*\to\widetilde{\nu}_{\tau}^\mathcal{R}\widetilde{\nu}_{\tau}^\mathcal{I})\
% \epsilon_{\text{sel}}^Z
% +\
% \sigma(pp\to W^*\to\widetilde{\nu}_{\tau}^\mathcal{R}\widetilde{\nu}_{\tau}^\mathcal{I}+\cdots)\
% \epsilon_{\rm sel}^W\right.
% \nonumber\\
% &&\left. +\sigma(pp\to \gamma^*,Z^*\to\widetilde{\nu}_{\tau}^\mathcal{R}\widetilde{\nu}_{\tau}^\mathcal{I}+\cdots)\
% \epsilon_{\text{sel}}^{\gamma ,Z}\right] \nonumber\\
% &\times&\left[\text{BR}(\widetilde{\nu}_{\tau}^\mathcal{R}\to\mu\mu/\mu e)\
% \epsilon_{\text{vert}}(c\tau^\mathcal{R})+\text{BR}(\widetilde{\nu}_{\tau}^\mathcal{I}\to\mu\mu/\mu
% e)\
% \epsilon_{\text{vert}}(c\tau^\mathcal{I})\right],
%  \label{numberevents}
% \end{eqnarray}
\begin{eqnarray}
{\#} \text{Dimuons} &=&
\left[\sigma(pp\to Z\to\widetilde{\nu}_{\tau}\widetilde{\nu}_{\tau})
\epsilon_{\text{sel}}^Z
+
\sigma(pp\to W\to\widetilde{\nu}_{\tau}\widetilde{\tau})
\epsilon_{\rm sel}^W
+\sigma(pp\to \gamma,Z\to\widetilde{\tau}\widetilde{\tau})
\epsilon_{\text{sel}}^{\gamma ,Z}
\right] 
\nonumber\\
&\times& 
\mathcal{L}\times \left[\text{BR}(\widetilde{\nu}_{\tau}^\mathcal{R}\to\mu\mu)\
\epsilon^{\mu\mu}_{\text{vert}}(c\tau^\mathcal{R})+\text{BR}
(\widetilde{\nu}_{\tau}^\mathcal{I}\to\mu\mu)\
\epsilon^{\mu\mu}_{\text{vert}}(c\tau^\mathcal{I})\right],
 \label{numberevents}
\end{eqnarray}
{where 
%$\text{BR}(\widetilde{\nu}_{\tau}\to\mu\mu)=
%\text{BR}(\widetilde{\nu}_{\tau}\to\tau\mu)\times 0.1739 
%+ \text{BR}(\widetilde{\nu}_{\tau}\to\tau\tau)\times (0.1739)^2 $ 
\begin{equation}
\text{BR}(\widetilde{\nu}_{\tau}\to\mu\mu)\equiv
\text{BR}(\widetilde{\nu}_{\tau}\to\tau\mu)\times 0.1739 
+ \text{BR}(\widetilde{\nu}_{\tau}\to\tau\tau)\times (0.1739)^2\,,
 \label{branching}
\end{equation}
with} 0.1739 the BR of the $\tau$ decay into muons (plus neutrinos), and we use an integrated luminosity of $\mathcal{L}= 20.3$ fb$^{-1}$~\cite{Aad:2015rba}
(300 fb$^{-1}$ when studying the 13-TeV prospects).
%
%where the selection efficiencies $\epsilon_{\rm sel}^Z$, $\epsilon_{\rm
%sel}^W$ and $\epsilon_{\rm sel}^{\gamma ,Z}$ are given {in 
%Tables~\ref{table:cutflow1}-\ref{table:cutflow2}.
%Tables~III and~V of Ref.~\cite{Lara:2018rwv}
%--\ref{table:cutflow2}.
The same formula can be applied for the other two channels.
%$e\mu$ channel shown in 
%Tables~IV and~VI, using the corresponding BRs, selection efficiencies, and vertex-level efficiencies (which turn out to be similar).}
%\ref{table:cutflow1}
%For the 13-TeV
%prospects the selection efficiencies {for the three channels} can be found in 
%Tables~VII--IX, and we use the same vertex-level efficiency {as in the 8-TeV case} and assume zero
%background}.
%\footnote{Notice that in the 13-TeV long-lived gluino search
%\cite{Aaboud:2017iio} the estimated number of background events is still
%much smaller than {one}, $\sim 10^{-2}$, which is similar in size to that
%in the 8-TeV search \cite{Aad:2015rba}. We therefore expect that the
%background in the 13-TeV dilepton displaced-vertex search is also as low
%as the 8-TeV one.} 
%As a result, 
If the predicted number of signal {events is above 3}
%the 2$\sigma$
%upper limit on visible number of events, 
the corresponding parameter
point of the model is excluded so that this is compatible with zero
number of events.

Let us finally remark that
in our analysis below, we scan the parameter space of the model and therefore 
$m_{\widetilde \nu_\tau}$ can be regarded as a continuous variable,
%~\footnote{ This is not strictly
%true since for discrete number of samples, there would be a discrete set of $m_{\widetilde \nu_\tau}$.
%The point is that the values of $m_{\widetilde \nu_\tau}$ for which the efficiencies were provided
unlike Ref~\cite{Lara:2018rwv} where the sneutrino masses used were 50, 60, 80 and 100 GeV.
%, however for us, $m_{\widetilde \nu_\tau}$
%can take any value between 50 and 100 GeV and can be regarded as a continuous variable.} 
%it would be important to take a pause and comment on how to proceed. 
%
%Concerning the vertex level efficiencies, namely $\epsilon_{\text{vert}}(c\tau)$, they have been
%obtained as the ratio of the number of signal events compatible with zero observed
%events, which in this case is 3, and that corresponding to the upper limits given in
%Ref.~\cite{Aad:2015rba}.
%As it can be noticed the $\epsilon_{\rm sel}$ and $\epsilon_{\text{vert}}(c\tau)$ shown so far
%as valid for for discrete and fixed set $m_{\widetilde \nu_\tau}$.
%Hence, in order to pass to a continuous limit of $m_{\widetilde \nu_\tau}$, $\epsilon_{\text{vert}}(c\tau)$ and $\epsilon_{\rm sel}$ are obtained by a polynomial fitting.
%Both, $\epsilon_{\text{vert}}(c\tau)$ and $\epsilon_{\rm sel}$ are obtained by a polynomial fit in function of $m_{\widetilde \nu_\tau}$.
For the selection efficiency we used a polynomial fitting from the discrete values of
$\epsilon_{\rm sel}$ given in Ref.~\cite{Lara:2018rwv} for each production mode, whereas for the vertex-level efficiency, the fitting
function is of the form $e^{P[\log(c\tau)]}$, where $P[x]$ is a polynomial in the variable $x$.

%%%%%%%%%%%%%%%%%%%%%%%%%%%%%%%%%%%%%%%%%%%%%%%%%%%%%%%%%%%%%%%%%%%%%%%%%%%%%%%%%%%%%%%%
\section{Strategy for the scanning 
%the tau left sneutrino LSP in the \texorpdfstring{$\mu\nu$SSM}{Lg}
} \label{methodology}

In this section we describe the methodology that we employed to search for points of our parameter space that 
are compatible with the current experimental data on neutrino and Higgs physics, as well as ensuring that the $\widetilde{\nu}_{\tau}$ is the LSP with a mass in the range of $45-100$ GeV.
In addition, we demanded
%via the likelihood 
the compatibility with some flavor observables.
To this end, we performed scans on the parameter space of the model, with the input parameters optimally chosen.
%to have a complete understanding 
%of 
%the phenomenology of a tau left sneutrino LSP, as well as that reproducing neutrino data.

%%%%%%%%%%%%%%%%%%%%%%%%%%%%%%%%%%%%%%%%%%%%%%%%
\subsection{Sampling the $\mn$}
%{Lg} 
%with a self developed code

%Bayesian inference methods have been successfully used in astronomical and cosmological
%data analysis, as well as in particle physics phenomenology.
%For example, using this method Ref.~\cite{Trotta:2008bp} analyzed the impact of the choice of priors and the influence
%of various constraints on the statistical conclusions for the preferred values of the parameters
%of the Constrained MSSM. The point is that with 

%Given the increasing amount of
%available experimental data, Bayesian inference methods appear very suitable in accessing viable
%regions of the parameter space of a given theoretical model.
For the sampling of the $\mn$, we used a likelihood data-driven method employing the
{\tt Multinest}~\cite{Feroz:2008xx} algorithm as optimizer. The goal is to find
regions of the parameter space of the $\mn$ that are compatible with a given experimental data. 

For it we have constructed the joint likelihood function: 
\begin{eqnarray}
 \mathcal{L}_{\text{tot}} =  \mathcal{L}_{\widetilde \nu_\tau} \times 
 \mathcal{L}_{\text{neutrino}} \times \mathcal{L}_{\text{Higgs}} \times \mathcal{L}_{\text{B physics}}
 \times \mathcal{L}_{\mu\text{ decay}} \times \mathcal{L}_{m_{\widetilde \chi^\pm}},
 \label{joint-likelihood}
\end{eqnarray}
where $\mathcal{L}_{\widetilde \nu_\tau}$ is basically the prior we impose on the tau left sneutrino mass, $\mathcal{L}_{\text{neutrino}}$ represents measurements of neutrino observables, $\mathcal{L}_{\text{Higgs}}$ Higgs observables, $\mathcal{L}_{\text{B physics}}$ B-physics constraints, $\mathcal{L}_{\mu\text{ decay}}$ $\mu$ decays constraints and 
$\mathcal{L}_{m_{\widetilde \chi^\pm}}$ LEPII constraints on the chargino mass.

%on the
%algorithm called {\tt Multinest}~\cite{Feroz:2008xx}.
%{\tt Multinest}
%uses a Bayesian inference method to 
%estimates a given
%set of parameters $\Theta$ in a given model $\mathcal{M}$ and for known data D, and relies on the Bayes theorem
%\begin{eqnarray}
% p(\Theta|\text{D},\mathcal{M}) = \frac{p(\text{D}|\Theta,\mathcal{M}) \times \mathcal{L}(\Theta)}{p(\text{D}|\mathcal{M})},
% \label{Bayes-theorem}
%\end{eqnarray}
%where $p(\text{D}|\Theta,\mathcal{M})$ is the prior distribution (before the data are seen),
%$\mathcal{L}(\Theta)$ the likelihood, $p(\text{D}|\mathcal{M})$ the Bayesian evidence,
%and $p(\Theta|\text{D},\mathcal{M})$ is the posterior (estimated parameters after data are seen) probability density. 
%More precisely, we developed a code that allows us to find
%regions of the parameter space of the $\mn$ that are compatible with a given experimental data. For each data,
%we defined a likelihood function and joined them in a combined likelihood function. 
%For a detailed description of MultiNest, the reader is referred to Ref.~\cite{Feroz:2008xx}.

%At first, {\tt Multinest} samples a set of active live points (2000 in our case) and for each point
%estimates the input parameters of the model also called priors.
%The number of priors, let say $n$, represents the dimension of the scan and
%will be labelled as $nD$ scan.
%Once the estimation is completed, the full set of parameters are then used as input for the spectrum calculation.
%(see Subsection ~\ref{spectrum-calculation}).
To compute the spectrum and the observables we used SARAH \cite{Staub:2013tta} to generate a 
%interfaced 
%{\tt Spheno} {{v}}3.3.6~\cite{Porod:2003um, Porod:2011nf} 
{\tt SPheno}~\cite{Porod:2003um, Porod:2011nf} version for the model. 
%with {\tt Multinest}.
%SPheno is a SUSY particles spectrum calculator program that computes (s)particles masses
%and mixing, decays widths, cross sections, BRs and low energy observables. 
%In order to make the scan faster in the computation of the spectrum, the gluino and squarks
%masses have been computed at tree level. This can be justified since the precision on their masses
%is small. For the rest of sparticles, except neutral scalars which receive full two loop corrections,
%the masses are computed at one loop. 
We condition that each point is required not to have tachyonic eigenstates. 
%BPs that do not to pass this constraint are discarded.
For the points that pass this constraint, we compute the likelihood associated to each experimental data set and 
%, as is discussed below. %Subsection~\ref{computation-of-likelihoods}.
for each sample all the likelihoods are collected in the joint likelihood
$\mathcal{L}_{\text{tot}}$ (see Eq.~(\ref{joint-likelihood}) above).

%%%%%%%%%%%%%%%%%%%%%%%%%%%%%%%%%%%%%%%%%%%%%
%\subsection{Computation of the spectrum} \label{spectrum-calculation}

%We interfaced SPheno~\cite{Porod:2003um, Porod:2011nf} with MultiNest to compute the spectrum.
%SPheno is a SUSY particles spectrum calculator program that computes (s)particles masses
%and mixing, decays widths, cross sections, branching ratios and low energy observables. Note that in order to make the scan faster in the computation of the spectrum, the gluino and squarks
%masses have been computed at tree level. This can be justified since the precision on their masses
%is small. For the rest of sparticles, except neutral scalars which receive full two loop corrections,
%the masses are computed at one loop. Note also that
%we require each sample to fulfill the physicality constraint consisting in checking the presence of tachyonic eigenstates. Benchmark points that do not to pass this constraint
%are discarded.

%%%%%%%%%%%%%%%%%%%%%%%%%%%%%%%%%%%%%%%%%%%
\subsection{Likelihoods}\label{computation-of-likelihoods}
We used three types of likelihood functions in our analysis. For observables
in which a measure is available we use a Gaussian likelihood function defined 
as follows
%for which the limit is provided as best fit and uncertainties,
%the likelihood function is a Gaussian,
\begin{eqnarray}
  \mathcal{L}(x) = \exp\left[-\frac{(x-x_0)^2}{2\sigma_T^2} \right],
  \label{eq:likelihood-Gaussian}
\end{eqnarray}
where $x_0 $ is the experimental best fit set on the parameter $x$, $\sigma_T^2= \sigma^2 +\tau^2$ with $\sigma$ and $\tau$ being respectively the experimental and theoretical uncertainties on the observable $x$.

On the other hand, for any observable for which the constraint is set as lower or upper limit, an example is the chargino mass lower bound, the likelihood function is defined as
\begin{eqnarray}
  \mathcal{L}(x)  &=& \frac{\sigma}{\sigma_T}  \left[ 1-K\left(D(x)\right) \right] \exp\left[-\frac{(x-x_0)^2p}{2\sigma_T^2} \right] 
  + \frac{1}{ \tau} K\left((x-x_0)p\right), 
    \label{eq:likelihood-lower+upper}
\end{eqnarray}
where
 %& \text{ where } & 
 \begin{eqnarray}
 D(x) = \frac{\sigma}{\tau} \left( \frac{(x_0 - x)p}{ \sigma_T} \right),
 \;
% \text{ and } 
 K(a) = \frac{1}{2} \text{erfc}\bigg(\frac{a}{\sqrt{2}}\bigg).
  \label{eq:likelihood-lower+upper2}
\end{eqnarray}
The variable $p$ takes $+1$ when $x_0$ represents the lower limit
and $-1$ in the case of upper limit, while \text{erfc} is the complementary error function.
%\begin{eqnarray}
% D(x) = \frac{\sigma}{\tau} \left( \frac{(x_0 - x)p}{ \sigma_T} \right)  , \quad 
% K(a) = \frac{1}{2} \text{erfc}\bigg(\frac{a}{\sqrt{2}}\bigg).
%\end{eqnarray} where
%\text{erfc} is the complementary error function.
%

The last class of likelihood function we used is a step function in such a way 
that the likelihood is one/zero if the constraint is satisfied/non-satisfied. 

%that is a 
%fixed $\chi^2$ value depending on whether the BP satisfies the limit or not. Note in this respect that the $\chi^2$ values are chosen to minimize the total $\chi^2_{\text{tot}}$ for viable points.

It is important to mention that in this work unless explicitly mentioned, the theoretical uncertainties $\tau$ are unknown and therefore are taken to be zero.
Subsequently, we present each constraint used in this work together with the corresponding type of likelihood function. 
%\begin{eqnarray}
% \mathcal{L}_{\text{tot}} =  \mathcal{L}_{\widetilde \nu_\tau} \times 
% \mathcal{L}_{\text{Neutrino}} \times \mathcal{L}_{\text{Higgs}} \times \mathcal{L}_{\text{B physics}}
% \times \mathcal{L}_{\mu\text{ decay}} \times \mathcal{L}_{m_{\widetilde \chi^\pm}}
% \label{joint-likelihood}
%\end{eqnarray}

%%%%%%%%%%%%%%%%%%%%%%%%%%%%%%%%%%%%%%%%
%\subsubsection{Tau left sneutrino mass}\label{section:tau-sneutrino-mass-length}

\vspace{0.5cm}

\noindent
{\bf Tau left sneutrino mass}

\noindent
In order to concentrate the sampling in the area in which the mass of the tau left
sneutrino $m_{\widetilde{\nu}_{\tau}} \in (45 , 100)$ GeV, we constructed a likelihood function $\mathcal{L}_{\widetilde \nu_\tau}$ which
is a Gaussian (see Eq.~(\ref{eq:likelihood-Gaussian})) with mean value 
$\mu_{m_{\widetilde{\nu}_{\tau}}} = 70$ GeV
%$\mu_{m_{\widetilde \nu}} = 70$ GeV
and width $\sigma_{m_{\widetilde{\nu}_{\tau}}} = 10$ GeV, and included it in the combined likelihood. 

%%%%%%%%%%%%%%%%%%%%%%%%%%%%%%%%%%%%%%%%
%\subsubsection{Neutrino observables}\label{section:neutrino-observables}

\vspace{0.5cm}

\noindent
{\bf Neutrino observables}

\noindent
%For sampling the $\mu\nu$SSM, 
We used the results for NO from Ref.~\cite{Capozzi:2017ipn} summarized in Table \ref{neutralinos-data-used},\footnote{{While we were doing the scan, we updated neutrino observables from a new neutrino global fit analysis \cite{Esteban:2018azc}.}}
where $\delta m^2=m_2^2-m_1^2$ and $\Delta m^2=m_3^2-(m_2^2+ m_1^2)/2$.
\begin{table}
    \centering
  \begin{tabular}{|c|c|c|c|c|c|}
  \hline %\toprule
  Parameters & $\sin^2{\theta_{12}}$& $\sin^2 {\theta_{13}}$ & $\sin^2 {\theta_{23}}$ &
  $ \delta m^2\, /\, 10^{-5}$  (eV$^2 $) & $\Delta m^2\, /\, 10^{-3}$ (eV$^2$) \\  \hline
  $\mu_{exp}$  & $0.297$ & $0.0215$ & $0.425$ &  $7.37$ & $2.525$ \\ 
  $\sigma_{exp}$  &  $0.017$ & $0.0007$ &  $0.021$ & $0.17$ & $0.042$  \\   
  \hline %  \bottomrule
  \end{tabular}
  \caption{ Neutrino data used in the sampling of the $\mu\nu$SSM.
  %, from Ref.~\cite{Capozzi:2017ipn}. 
  %The values are taken from  \cite{Capozzi:2017ipn}.
  }
\label{neutralinos-data-used}
\end{table}
For each of the observables listed in the neutrino sector, the likelihood function is
a Gaussian (see Eq.~(\ref{eq:likelihood-Gaussian})) centered at the mean value $\mu_{exp}$ and with width $\sigma_{exp}$. {Concerning the cosmological upper
bound on the sum of the masses of the light active neutrinos given
by $\sum m_{\nu_i} < 0.12$ eV \cite{Aghanim:2018eyx}, even though we did not include it directly in the total likelihood, we imposed it on the viable points obtained.}
%, and collected in $\mathcal{L}_{\text{neutrino}}$.

%%%%%%%%%%%%%%%%%%%%%%%%%%%%%%%%%%%%%%%%%%%
%\subsubsection{Higgs observables}\label{section:Higgs-observables}

\vspace{0.5cm}

\noindent
{\bf Higgs observables}

\noindent
Before the discovery of the SM-like Higgs boson, the negative searches of Higgs signals at the Tevatron, LEP and LHC, were transformed into exclusions limits that must
be used to constrain any model. Its discovery at the LHC added crucial constraints that
must be taken into account in those exclusion limits. 
We have considered all these constraints in the analysis of the $\mn$, where the Higgs sector is extended with respect to the MSSM as discussed in Section~\ref{section0}.
For constraining the predictions in that sector of the model, we interfaced 
{\tt HiggsBounds} {{v}}5.3.2~\cite{Bechtle:2008jh,Bechtle:2013wla} 
%{\tt HiggsBounds} {{v}}4.3.1~\cite{Bechtle:2008jh,Bechtle:2013wla}  
with MultiNest.
First, several theoretical predictions in the Higgs sector (using a $\pm 3$ GeV theoretical uncertainty on the SM-like Higgs boson) are provided to determine which process has the highest exclusion power, according to the
list of expected limits from LEP and Tevatron. Once the process with the highest statistical
sensitivity is identified, the predicted production cross section
of scalars and pseudoscalars multiplied by the BRs are compared with the limits set by
these experiments. Then, whether the corresponding point of the
parameter under consideration is allowed or not at 95\% confidence level is indicated.
In constructing the likelihood from HiggsBounds constraints, the likelihood function is taken to be a step function. Namely, it is set to one for points for which Higgs physics is realized, and zero otherwise. 
%This choice allows to minimize the $\chi^2$ for points that are allowed while penalizing the ones that are excluded.
%As already mentioned HiggsBounds algorithm does not offer the possibility of verifying whether
%a given Higgs scalar of a given model is in agreement with the signal that has been observed at
%CMS and ATLAS. 
Finally, in order to address whether a given Higgs scalar of the $\mn$ 
is in agreement with the signal observed by ATLAS and CMS, we interfaced 
{\tt HiggsSignals} {{v}}2.2.3~\cite{Bechtle:2015pma,Bechtle:2013xfa} 
%{\tt HiggsSignals} {{v}}1.4.0~\cite{Bechtle:2015pma,Bechtle:2013xfa} 
with MultiNest.
%to test the model prediction against the measured mass and signal strength discovered by ATLAS and
%CMS collaborations. 
A $\chi^2$ measure is used to quantitatively determine the compatibility of the $\mn$ prediction with the measured signal strength and mass. 
The experimental data used are those of the LHC with some complements from Tevatron. The details of the likelihood evaluation can be found in Refs.~\cite{Bechtle:2015pma,Bechtle:2013xfa}. 
%We denote by $ \mathcal{L}_{\text{Higgs}}$ the likelihood associated to Higgs observables.

%%%%%%%%%%%%%%%%%%%%%%%%%%%%%%%%%%%%%%%%%%%%%

\vspace{0.5cm}

\noindent
{\bf B decays}

\noindent
$b \to s \gamma$ is a flavour changing neutral current (FCNC) process, and hence it is forbidden
at tree level in the SM. However, its occurs at leading order through loop diagrams.
Thus, the effects of new physics (in the loops) on the rate of this 
process can
be constrained by precision measurements. 
In the combined likelihood, we used the average value of $(3.55 \pm 0.24) \times 10^{-4}$ provided in Ref.~\cite{Amhis:2012bh}. Notice that the likelihood function is also a Gaussian
(see Eq.~(\ref{eq:likelihood-Gaussian})).
%%%%%%%%%%%%%%%%%%%%%%%%%%%%%%%%%%%%%%%%%%%%%%
Similarly to the previous process, $B_s \to \mu^+\mu^-$ and  $B_d \to \mu^+\mu^-$
%$b \to s \gamma$ process, its decays into a pair of muons 
are also forbidden at tree level in the SM but occur radiatively.
In the likelihood for these observables (\ref{eq:likelihood-Gaussian}), we used the combined results of LHCb and CMS~\cite{CMSandLHCbCollaborations:2013pla}, 
$ \text{BR} (B_s \to \mu^+ \mu^-) = (2.9 \pm 0.7) \times 10^{-9}$ and
$ \text{BR} (B_d \to \mu^+ \mu^-) = (3.6 \pm 1.6) \times 10^{-10}$. 
Concerning the theoretical uncertainties for each of these observables we take $\tau= 10 \%$ of the corresponding best fit value.
We denote by $\mathcal{L}_{\text{B physics}}$
the likelihood from $b \to s \gamma$, $B_s \to \mu^+\mu^-$ and $B_d \to \mu^+\mu^-$.

% \times \mathcal{L}_{\mu\text{ decay}} \times \mathcal{L}_{m_{\widetilde \chi^\pm}}

%%%%%%%%%%%%%%%%%%%%%%%%%%%%%%%%%%%%%%%%%%%%%%%%
%\subsubsection{  \texorpdfstring{ $\mu \to e \gamma$ and $\mu \to e e e$}{Lg} } \label{muTo3e+egamma}

\vspace{0.5cm}

\noindent
{\bf $\mu \to e \gamma$ and $\mu \to e e e$}

\noindent
We also included in the joint likelihood the constraint from 
BR$(\mu \to e\gamma) < 5.7\times 10^{-13}$ and BR$(\mu \to eee) < 1.0 \times 10^{-12}$.
For each of these observables we defined the likelihood as a step function. As explained before, if a point is in agreement with the data, the likelihood
$\mathcal{L}_{\mu\text{ decay}}$ is set to 1 otherwise to 0.

{Let us point out here that we did not try to 
explain the interesting but not conclusive 3.5$\sigma$ discrepancy between the measurement of the
anomalous magnetic moment of the muon and the SM prediction, 
$\Delta a_{\mu}= a_{\mu}^{\text{exp}}-a_{\mu}^{\text{SM}}= (26.8 \pm 6.3\pm 4.3) \times 10^{-10} $ \cite{Tanabashi:2018oca}.
Since we decouple the rest of the SUSY spectrum with respect to the tau left sneutrino mass, we do not expect a large SUSY contribution over the SM value.
%$a^{\text{exp}}_\mu =11659209.1 \pm 5.4 \pm 3.3 \times 10^{-10}$ \cite{Tanabashi:2018oca}. 
We checked for the points fulfilling all constrains discussed in Section~\ref{results-scans},
that the extra contribution $a_{\mu}^{\text{SUSY}}$ is within the SM uncertainty.}

%%%%%%%%%%%%%%%%%%%%%%%%%%%%%%%%%%%%%%%%%%
%\subsubsection{Chargino mass bound}\label{chargino-mass-bound}
%Before closing the paragraph on the constraints used in the joint likelihood, we would like to
%comment that 

\vspace{0.5cm}

\noindent
{\bf Chargino mass bound}

\noindent
In RPC SUSY,
%(e.g. MSSM, NMSSM)
%it is often used
%widely assumed that 
%for 
the lower bound on the lightest chargino mass of about $94$ GeV
%from LEP searches. Note however that the lower limit actually 
depends on the spectrum of the model~\cite{Tanabashi:2018oca,Sirunyan:2018ubx}.
Although in the $\mn$ there is RPV and therefore this constraint does not apply automatically, to compute $\mathcal{L}_{m_{\widetilde \chi^\pm}}$ we have chosen a conservative limit of $m_{\widetilde \chi^\pm_1} > 92$
GeV with the theoretical uncertainty $\tau= 5 \%$ 
%$\tau = 5 \%$ 
of the chargino mass.

%\vspace{0.5cm}

%\noindent
%In sum, the likelihood function that we used to find viable points compatible with
%the experimental data, is 
%\begin{eqnarray}
% \mathcal{L}_{\text{tot}} =  \mathcal{L}_{\widetilde \nu_\tau} \times 
% \mathcal{L}_{\text{neutrino}} \times \mathcal{L}_{\text{Higgs}} \times \mathcal{L}_{\text{B physics}}
% \times \mathcal{L}_{\mu\text{ decay}} \times \mathcal{L}_{m_{\widetilde \chi^\pm}}.
% \label{joint-likelihood}
%\end{eqnarray}
%where $\mathcal{L}_{\widetilde \nu_\tau} $ corresponds to the constraints on $m_{\widetilde \nu_\tau}$
%described (section~\ref{section:tau-sneutrino-mass-length}) and
%$ \mathcal{L}_{\text{Neutrino}} $ for neutrino physics data (section~\ref{section:neutrino-observables}).
%$ \mathcal{L}_{\text{Higgs}} $ refers to the constraints implemented in HiggsBounds and HiggsSignals
%and discussed (section~\ref{section:Higgs-observables}). Concerning 
%$\mathcal{L}_{\text{B physics}} $ it represents the chi square from $b\to s\gamma$, $B_s, B_d \to \mu\mu$ 
%(sections~\ref{bTosgamma} and \ref{BsdTomumu}). 
%$\mathcal{L}_{\mu\text{ decay}} $ and $\mathcal{L}_{m_{\widetilde \chi^\pm}} $ stand respectively for the
%constraints from $\mu \to e e e, e\gamma$ (section~\ref{muTo3e+egamma}) and for
%lower chargino mass bound (section~\ref{chargino-mass-bound}).

%%%%%%%%%%%%%%%%%%%%%%%%%%%%%%%%%%%%%%%%%%%%%%%
\subsection{Input parameters}
\label{choice-of-input-for-scan}
\label{input}

%Given the limitations in computing resources,
In order to efficiently scan for the 
$\widetilde{\nu}_{\tau}$ LSP in the $\mu\nu$SSM with a mass in the range $45-100$ GeV,
it is important to identify first
the 
%most relevant 
parameters to be used, and optimize their number and their ranges of values.
This is what we carry out here, where we discuss the most relevant parameters for obtaining correct neutrino and Higgs physics,
providing at the same time the 
$\widetilde{\nu}_{\tau}$ as the LSP with the mass 
%\R{and decay length} 
in the desired range.
%, \R{and with the right phenomenology}.

%As discussed in Eq.~(\ref{freeparameters}), 
The relevant parameters in the neutrino sector of the $\mn$ are 
$\lambda, \kappa, v_R, v_i, Y_{\nu_i}, \tan\beta$ and $M$ (see Eq.~(\ref{freeparameters})).
Since $\lambda, \kappa$ and  $v_R$ are crucial for Higgs physics, we will fix first them to appropriate values. The parameter
$\tan\beta$ 
is also important for both, Higgs and neutrino physics, thus we will consider a narrow range of possible values to ensure good Higgs physics.
Concerning
$M$, which is a kind of average of bino and wino soft masses (see Eq.~(\ref{effectivegauginomass2})), inspired by GUTs we will assume $M_2 = 2M_1$, and scan over $M_2$.
On the other hand, 
sneutrino masses introduce in addition the parameters 
$T_{{\nu}_i}$ (see Eq.~(\ref{evenLLL2})). In particular, $T_{{\nu}_3}$ is the most relevant one for our discussion of the $\widetilde{\nu}_{\tau}$ LSP, and we will scan it in an appropriate range of small values. Since the left sneutrinos of the first two generations must be heavier, we will fix $T_{{\nu}_{1,2}}$ to a larger value.

Summarizing, we will perform scans over the 9 parameters 
$Y_{\nu_i}, v_i, T_{{\nu}_3}, \tan\beta, M_2$, as shown in Table~\ref{Scans-priors-parameters},
using log priors (in logarithmic scale) for all of them, except for
$\tan\beta$ which is taken to be a flat prior (in linear scale).
The ranges of $v_i$ and $Y_{\nu_i}$ are natural in the context of the electroweak-scale seesaw of the $\mn$.
%Considering those values for the neutrino Yukawa couplings, 
The range
of $T_{\nu_{3}}$ is also natural if we follow the usual assumption based on the supergravity framework discussed in Eq.~(\ref{evenLLL22}) that the trilinear parameters are proportional to the corresponding Yukawa couplings, i.e. in this case $T_{\nu_{3}}= A_{\nu_{3}} Y_{\nu_3}$ implying 
$-A_{\nu_{3}}\in$ ($1, 10^{4}$) GeV.
%Note that under this assumption, Eq.~(\ref{evenLLL2}) can be written for the tau left sneutrino as
%\bea
%m_{\widetilde{\nu}_{\tau}}^2
%\approx  
%\frac{Y_{{\nu}_3}v_u}{2v_{3}}v_{R}
%\left(
%-\sqrt 2 A_{{\nu}_3}-\kappa v_{R}
%+
%\frac{3\lambda v_{R}}{\tan\beta}
%\right).
%\label{evenLLL21}
%\eea
Concerning $M_2$, its range of values is taken such that a bino at the bottom
of the neutralino spectrum leaves room to accommodate a 
$\widetilde{\nu}_{\tau}$ LSP with a mass below 100 GeV. 
Scans 1 ($S_1$) and 2 ($S_2$) correspond  
to different values of $\tan\beta$, and other benchmark
parameters as shown in 
Table~\ref{Scans-fixed-parameters}.

{In 
Table~\ref{Scans-fixed-parameters} we choose first two values of $\lambda$, covering a representative region of this parameter.
From
a small/moderate value, $\lambda\approx 0.1$ ($S_1$), to a large value, $\lambda\approx 0.4$  ($S_2$), in the border of perturbativity up to the GUT scale~\cite{Escudero:2008jg}.
For scan $S_1$, since $\lambda$ is small we are in a similar situation as in the MSSM, and moderate/large values of $\tan\beta$, $|T_{u_{3}}|$, and soft stop masses, are necessary to obtain the correct SM-like Higgs mass. 
In addition, if we want to avoid the chargino mass bound of RPC SUSY, 
the value of $\lambda$ also force us to choose a moderate/large value of $v_R$ to obtain a large enough value of 
$\mu=3 \lambda \frac{v_{R}}{\sqrt 2}$.
In particular, we choose $v_R=1750$ GeV giving rise to $\mu\approx 379$ GeV.
The latter parameters, 
$\lambda$ and $v_R$, together with $\kappa$ and $T_{\lambda}$ are also relevant to obtain the correct values of the off-diagonal terms of the mass matrix mixing the right sneutrinos with Higgses.
As explained in Eq.~(\ref{sps-approx2}), the parameters $\ka$ and $v_R$ (together with
$T_{\kappa}$) are also crucial to determine
the mass scale of the right sneutrinos.
In scan $S_1$, where we choose $T_{\kappa}=-390$ GeV to have heavy pseudoscalar right sneutrinos (of about 1190 GeV), the value of $\kappa$ has to be large enough in order to avoid 
too light (even tachyonic) scalar right sneutrinos. Choosing $\kappa=0.4$, we get masses for the latter of about $700-755$ GeV.
}

\begin{table}
\begin{center}
\begin{tabular}{|l|l|}
\hline
 \multicolumn{1}{|l|}{\bf Scan 1 ($S_1$)}&\multicolumn{1}{l|}{ \bf Scan 2 ($S_2$)}\\
\cline{1-2}
  $\tan\beta \in (10, 16)$ & $\tan\beta \in (1, 4)$\\ 
\hline
\multicolumn{2}{|l|}{ \quad \quad $Y_{\nu_{i}} \in (10^{-8} , 10^{-6})$ }\\
\multicolumn{2}{|l|}{ \quad \quad $v_i \in (10^{-6} , 10^{-3})$  }\\
\multicolumn{2}{|l|}{ \quad \quad  $-T_{\nu_{3}} \in (10^{-6} , 10^{-4})$ }\\
\multicolumn{2}{|l|}{ \quad \quad $M_2 \in (150 , 2000)$ }\\
\cline{1-2}
\end{tabular}
\end{center}
  \caption{Range of low-energy values of the input parameters that are varied in the two scans, where  $Y_{\nu_{i}}$, $v_i$, $T_{\nu_{3}}$ and $M_2$ are $\log$ priors while
$\tan\beta$ is a flat prior. 
%We assume $M_2=2M_1$. 
The
VEVs $v_i$, and the soft parameters $T_{\nu_{3}}$ and $M_2$, are given in GeV.}
 \label{Scans-priors-parameters}
\end{table} 

\begin{table}
\begin{center}
\begin{tabular}{|c|c|c|}
    \hline
     {\bf Parameter}&  {\bf Scan 1 ($S_1$) } &  {\bf Scan 2 ($S_2$) }\\   
    \hline 
    $\lambda$     & 0.102  &  0.42  \\ 
     \hline   
    $\kappa$      & 0.4   &   0.46 \\ 
    \hline
    $v_R$     & 1750 &  421  { }    \\ 
    \hline
    $T_{\lambda}$   & 340   &  350  \\ 
    \hline
    $-T_{\kappa}$    & $390$  &  $108$ \\ 
    \hline
    $-T_{u_{3}}$  & $4140$  & $1030$ \\ 
    \hline
    $m_{\widetilde Q_{3L}}$  & 2950 & 1972 \\ 
    \hline
    $m_{\widetilde u_{3R}}$   & 1140 & 1972 \\ 
    \hline
    $ M_3$  & \multicolumn{2}{|c|}{ 2700 }\\
    \hline  
    $m_{\widetilde Q_{1,2L}}, m_{\widetilde u_{1,2R}}, 
    m_{\widetilde d_{1,2,3R}}, m_{\widetilde e_{1,2,3R}}$ & \multicolumn{2}{|c|}{ 1000 }\\
    \hline
    $T_{u_{1,2}}$  & \multicolumn{2}{|c|}{ 0}\\
    \hline
    $T_{d_{1,2}}$, $T_{d_{3}}$  & \multicolumn{2}{|c|}{ 0, $100$ }\\
    \hline    
    $T_{e_{1,2}}$, $T_{e_{3}}$  & \multicolumn{2}{|c|}{ 0, $40$ }\\
    \hline
    $-T_{\nu_{1,2}}$   & \multicolumn{2}{|c|}{ $10^{-3}$ }  \\
    \hline 
    \end{tabular}  
\end{center}
   \caption{Low-energy values of the input parameters that are fixed in the two scans.
The VEV $v_R$ and the soft trilinear parameters, soft gluino masses and soft scalar masses
%$T$'s, $M_3$ and $m$'s below, and 
are given in GeV.
%    $\lambda_i =\lambda$, $\kappa_{iii} =\kappa$ and 0 otherwise, $T_{\lambda_i} =T_\lambda$,
%    $T{\kappa_{iii}} =T_\kappa$ and 0 otherwise and $v_{R_i} = v_R$ 
}
     \label{Scans-fixed-parameters}
\end{table}

{For scan $S_2$, where we choose a large value for $\lambda$, we are in a similar situation as in the NMSSM, and
a small value of $\tan\beta$, and moderate values of $|T_{u_{3}}|$ and soft stop masses,
are sufficient to reproduce the correct SM-like Higgs mass.
Now, a moderate value of $v_R$ is sufficient to obtain a large enough value of $\mu$. In particular, we choose $v_R=421$ GeV giving rise to $\mu\approx 375$ GeV.
This value of $v_R$ implies that $|T_{\kappa}|$ cannot be as large as for scan $S_1$
because then a too large value of $\kappa$ would be needed to avoid tachyonic scalar
right sneutrinos. Thus we choose $T_{\kappa}=-108$ GeV, and $\kappa=0.42$, which produces scalar and pseudoscalar sneutrinos lighter than in scan $S_1$ but still heavier than
$\widetilde{\nu}_{\tau}$ LSP and left stau NLSP.
In particular, their masses are in the ranges $225-256$ GeV and $345-355$ GeV, respectively.}

{The values of the parameters shown below 
$m_{\widetilde u_{3R}}$ in Table~\ref{Scans-fixed-parameters},
concerning gluino, and squark and slepton masses, and quark and lepton trilinear parameters, are not specially relevant for our analysis, and
we choose for each of them the same values for both scans. 
Finally, compared to the values of $T_{\nu_{3}}$, the values chosen for
$T_{\nu_{1,2}}$ are natural within our framework
$T_{\nu_{1,2}}= A_{\nu_{1,2}} Y_{\nu_{1,2}}$, since
larger values of the Yukawa couplings are required for similar values of 
$A_{\nu_{i}}$.
%These scans will allow us to compare our results with those obtained in Ref.~\cite{Lara:2018rwv}.
%%
In the same way, the values of $T_{d_3}$ and $T_{e_3}$ have been chosen taking into account the corresponding Yukawa couplings}

%%%%%%%%%%%%%%%%%%%%%%%%%%%%%%%%%%%%%%%%%%%%%%%%%%%%%%%%%%%%%%%%%%%%%%%%%%%%%%%%%%%%%%%%
\section{Results}
\label{results-scans}

By using the methods described in the previous sections, we
evaluate now the current and potential limits on the parameter space of our scenario from the displaced-vertex searches with the 8-TeV
ATLAS result~\cite{Aad:2015rba}, and discuss the prospects for the 13-TeV searches.

To find regions consistent with experimental observations we have performed about 72 million of spectrum evaluations in total and the total amount of computer required for this was approximately 380 CPU years. 

To carry this analysis out, we follow several steps. First, we select points from the scan that lie within $\pm 3\sigma$ of all
neutrino physics observables, namely the mixing angles and mass squared differences. Second, we
put  $\pm 3\sigma$ cuts from $b \to s \gamma$, $B_s \to \mu^+\mu^-$ and $B_d \to \mu^+\mu^-$.
The points that pass these cuts are required to satisfy also the upper limits of $\mu \to e \gamma$ and $\mu \to eee$. 
The third step in the selection of our points is to ensure 
a tau left sneutrino LSP with $m_{\widetilde \nu_\tau} \in (45, {100})$ GeV,
{and the left stau as the NLSP}.
In the fourth step we impose 
that Higgs physics is realized. As already
mentioned, we use {\tt HiggsBounds} and
{\tt HiggsSignals} taking into account
the constraints from the latest 13-TeV results. In particular, we require 
that the p-value reported by {\tt HiggsSignals} be larger than 5 \%.
{It is worth noticing here that, with the help of 
Vevacious~\cite{Camargo-Molina:2013qva}, we have also checked that the
EWSB vacua corresponding to the
previous allowed points are stable.}

The final set of cuts is related to $\widetilde{\nu}_{\tau}$ LSP
searches with displaced vertices. 
From the points left above, we select those with 
%$m_{\widetilde \nu_\tau} \in (50, 100)$ GeV and 
decay length $c\tau > 0.1$ mm in order to be constrained by the current experimental results, as mentioned in previous sections.
%~\ref{section2}. 
%\R{Of the remaining, we make sure to remove those for which $\widetilde \tau$ is not the NLSP.}
Finally, since the number of signal events compatible with zero observed
events is 3, we look for points with a number of signal events {above 3}.
%satisfying $N_{\text{events}} < 3$. 

%In Subsection~\ref{neutrinos}, 

\begin{figure}[t]
 \centering
\includegraphics[width=\linewidth, height=0.5\textheight]{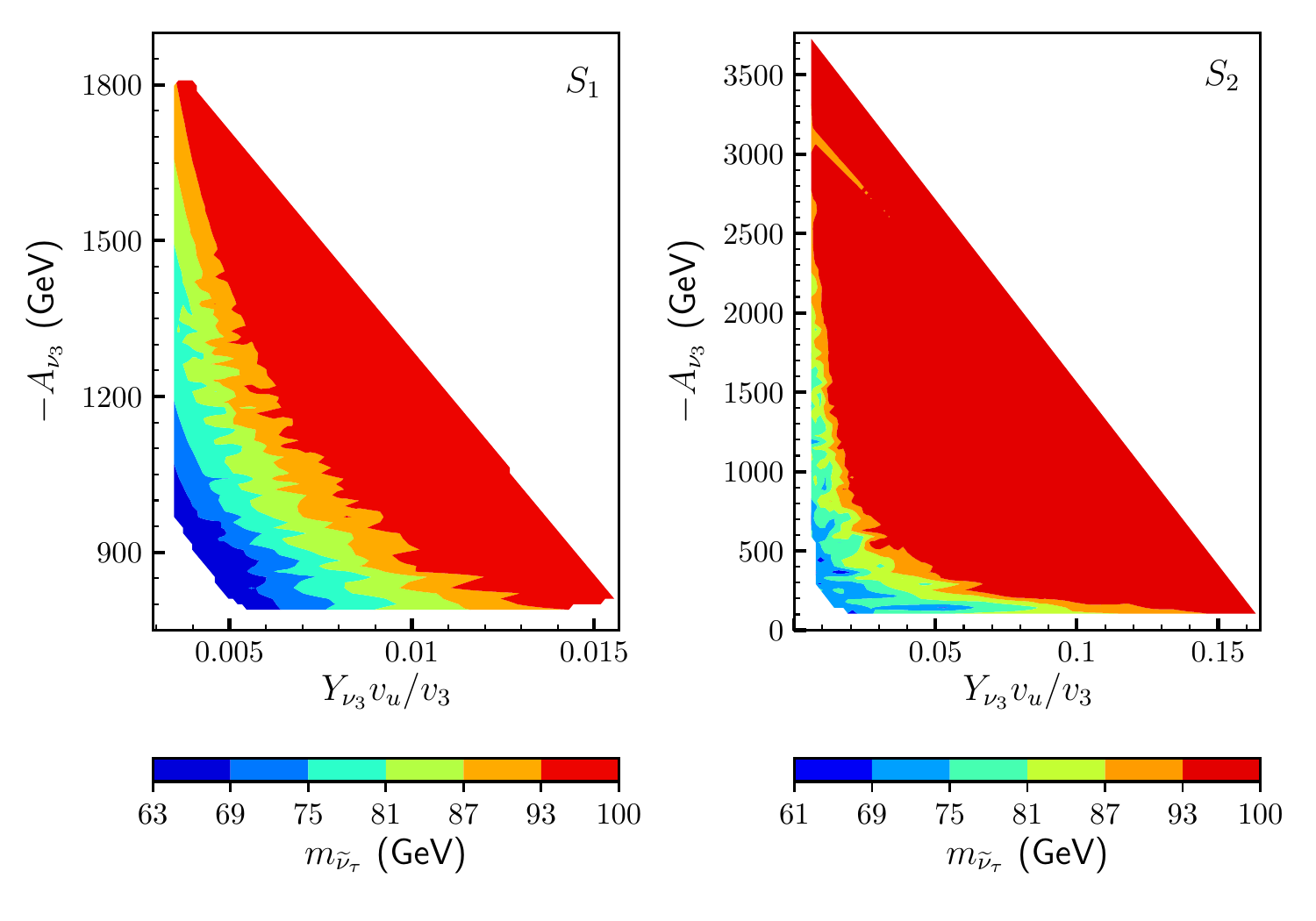}
 \caption{$-A_{\nu_3}$ versus 
$Y_{\nu_3} v_u/{v_3}$ for scan $S_1$ (left) and scan $S_2$ (right).
 The colours indicate different values of the tau left sneutrino LSP mass.}
 \label{S1S2-2D-AvL3-Prefact-MSvL3.png}
\end{figure}

\subsection{Constraints from neutrino/sneutrino physics.}
\label{neutrinos}

As discussed in detail in Section~\ref{section0}, reproducing neutrino physics is an important asset of the 
$\mn$. It is therefore important to analyze first 
the constraints imposed by this requirement on the relevant parameter space of the model when the
$\widetilde{\nu}_{\tau}$ is the LSP.

Imposing all the cuts discussed above, with the exception of the one associated to the number of signal events,
we show in Fig.~\ref{S1S2-2D-AvL3-Prefact-MSvL3.png} the values of the parameter 
$A_{\nu_3}$ versus the prefactor in Eq.~(\ref{evenLLL22}),
$Y_{\nu_3} v_u/{v_3}$, giving rise to a mass of the $\widetilde{\nu}_{\tau}$ in the desired range $45-100$ GeV. The colours indicate different values
of this mass.
Scan $S_1$ ($S_2$) is shown in the left (right)-hand side of the figure.
Let us remark that these plots 
have been obtained using the full numerical computation including loop 
corrections, 
although the tree-level mass in Eq.~(\ref{evenLLL22}) gives a good qualitative idea of the results.
In particular, in scan $S_1$ we can see that the allowed range of $-A_{\nu_3}$ is $779-1820$ GeV, corresponding to
$-T_{\nu_3}$ in the range 
$8.3\times 10^{-6}-3.5\times 10^{-5}$ GeV.
We can also see, as can be deduced from Eq.~(\ref{evenLLL22}), that 
for a fixed value of $-A_{\nu_3}$ ($Y_{\nu_3} v_u/{v_3}$) the greater $Y_{\nu_3} v_u/{v_3}$ ($-A_{\nu_3}$) is, the greater 
$m_{\widetilde{\nu}_{\tau}}$ {becomes}. 
For scan $S_2$, the allowed range 
of $-A_{\nu_3}$ turns out to be $67-3764$ GeV,
corresponding to
$-T_{\nu_3}$ in the range 
$2.1\times 10^{-6}-4.9\times 10^{-5}$ GeV.
The differences in the range of allowed values for $A_{\nu_3}$ and $Y_{\nu_3} v_u/{v_3}$
of the scan $S_1$ with respect to $S_2$, are due to the negative vs. the positive contribution of the sum of the second and third terms in the bracket of Eq.~(\ref{evenLLL22}), respectively, as well as to the different values of $v_R$ which appears also as a prefactor in that equation.

\begin{figure}[t!]
 \centering
\includegraphics[width=\linewidth, height=0.55\textheight]{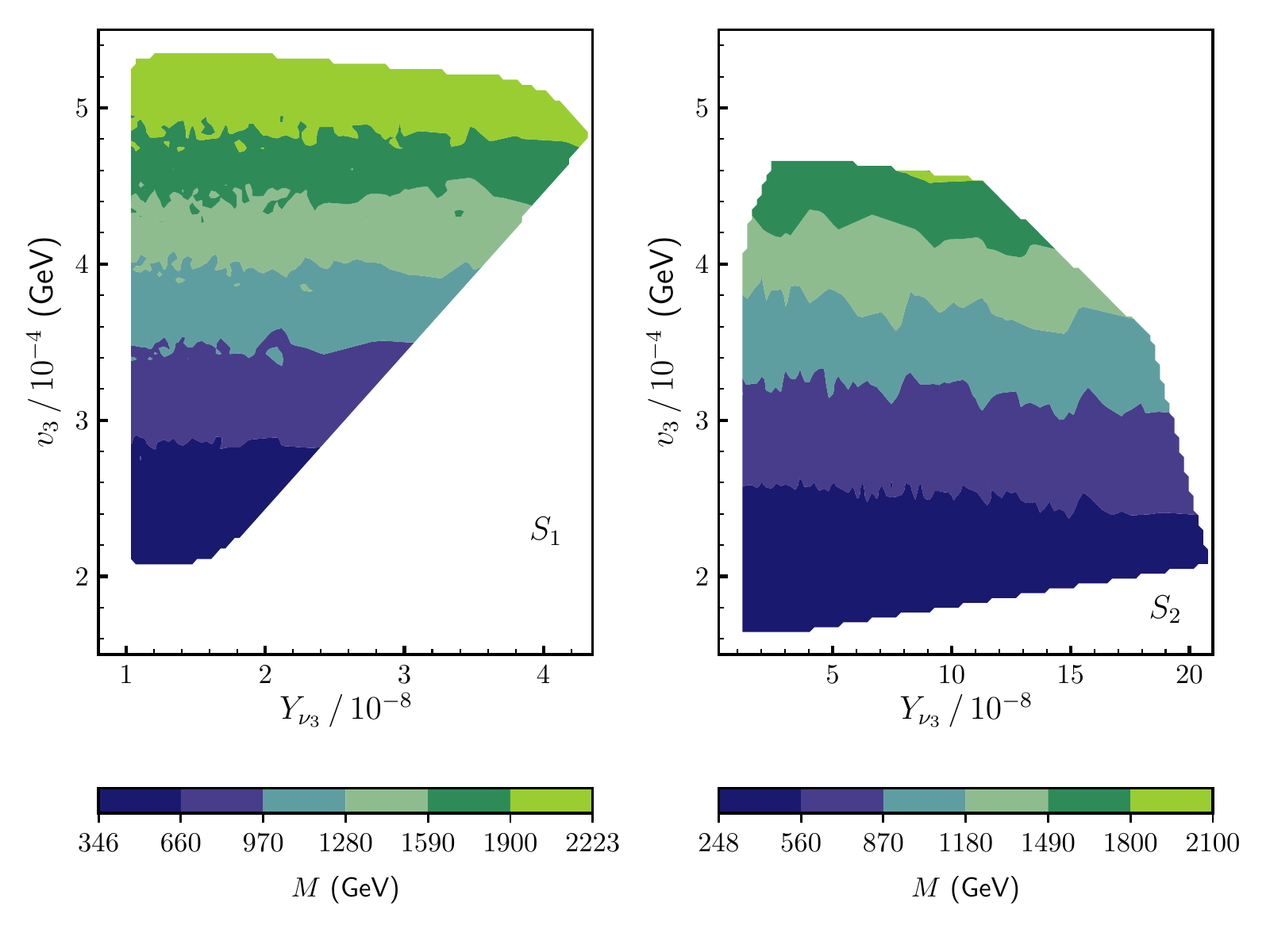}
 \caption{$v_3$ versus 
$Y_{\nu_3}$ for scan $S_1$ (left) and scan $S_2$ (right).
 The colours indicate different values of the gaugino mass parameter $M$ defined
 in Eq.~(\ref{effectivegauginomass2}).}
 \label{S1S2-2D-MSvL3-Params.png}
\end{figure}

\begin{figure}[t!]
  \centering
\includegraphics[width=\linewidth, height=0.55\textheight]{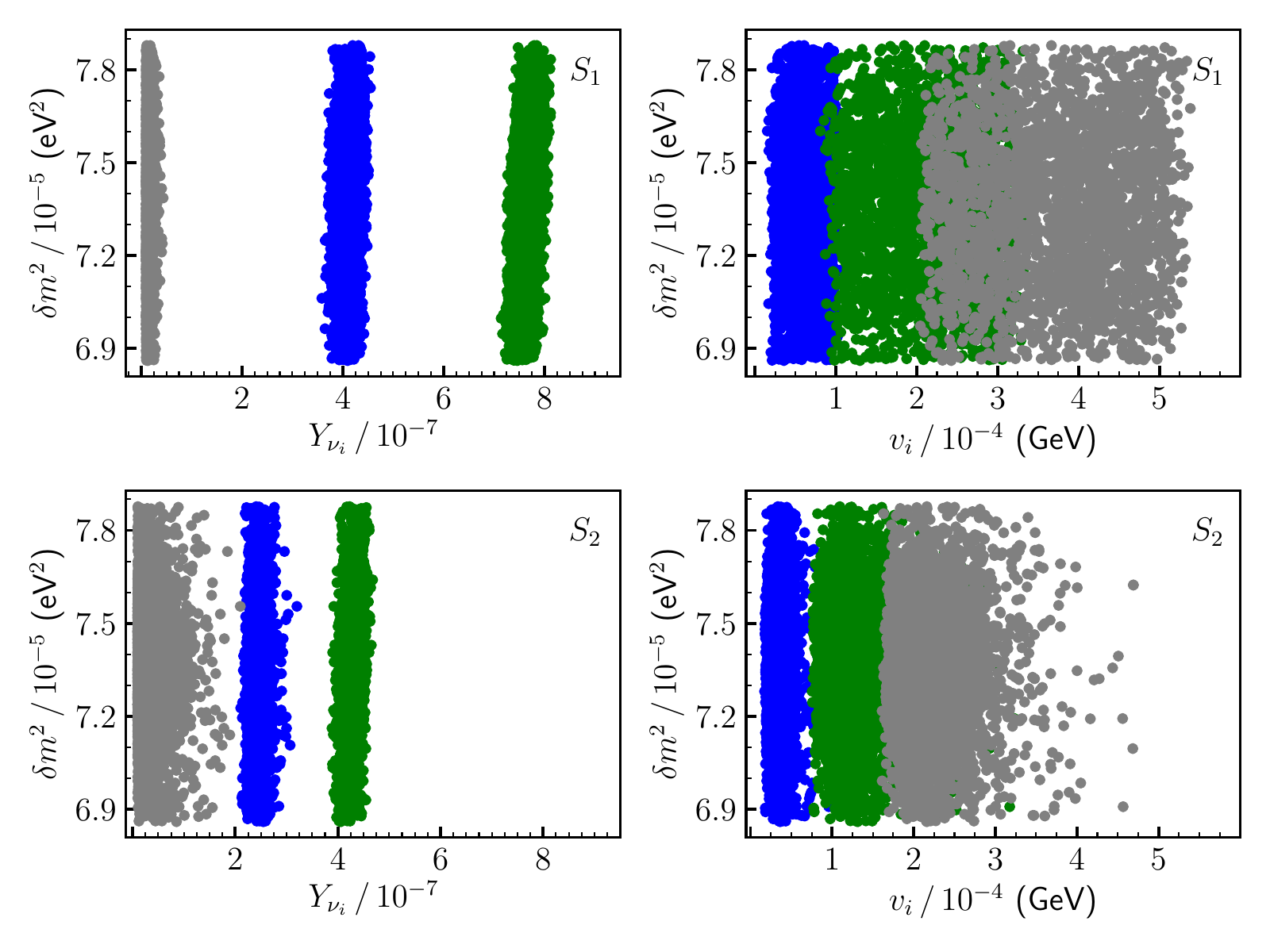}
\caption{$\delta m^2$ versus neutrino Yukawas (left) and left sneutrino VEVs (right) for scan $S_1$ (top) and
 $S_2$ (bottom).
 Colors blue, green and grey correspond to $i=1,2,3$,
 %the third, first and second family,
 respectively.
}
 \label{S1S2-NuParams-vs-Yvi.png}
\end{figure}

Let us finally note that $m_{\widetilde \nu_{\tau}}$ is always larger than about {61 GeV},
which corresponds to half of the mass of the SM-like Higgs (remember that we allow a $\pm$3 GeV theoretical uncertainty on its mass).
For smaller masses, the latter would dominantly decay into sneutrino pairs, {leading to an inconsistency with Higgs data}.\footnote{{In this scenario 
the SM-like Higgs 
decays into pairs of scalar/pseudoscalar tau left sneutrinos via gauge interactions, 
mostly from D-terms $\sim \frac{1}{4}(g^2+g'^2)\widetilde \nu_i \widetilde \nu_i^*  H^0_uH^{0*}_u$,
since its largest component is $H^0_u$}.}
%We also note that $m_{\widetilde \nu_{\tau}}$ increases in scan $S_1$ with larger (smaller) values of $|T_{\nu_3}|$ ($Y_{\nu_3}$).
%This can be easily understood from
%Eq.~(\ref{evenLLL2}), taking into account that
%the sum of the two terms in the parenthesis is basically fixed around the negative value $-660$ GeV.
%For scan $S_2$,
%the variation in $Y_{\nu_3}$ is not so relevant
%because this sum varies in a large range between 340 and $-60$ GeV, due to the variation in $\tan\beta$.
%Since the VEV $v_3$ also enters in the tree-level formula for $m_{\widetilde \nu_{\tau}}$,
%we also show in Figs.~\ref{S1-2D-MSvL3-Params.png} and~\ref{S2-2D-MSvL3-Params.png} this mass but in the plane
%$-T_{\nu_{3}}$ vs. $v_3$. 

In Fig.~\ref{S1S2-2D-MSvL3-Params.png}, we show 
$v_3$ vs. $Y_{\nu_3}$ for scan $S_1$ (left) and scan $S_2$ (right), with
the colours indicating now different values of $M$.
There we can see that the greater $v_3$ is, the greater $M$ becomes.
In addition, for a fixed value of $v_3$, $M$ is quite
independent of the variation in $Y_{\nu_3}$. This confirms that,
as explained in solution 2) of Subsection~\ref{neusneu}, the gaugino seesaw is
the dominant one for the third neutrino family.
From the figure, we can see that the range of $M$ reproducing the
correct neutrino physics is $346-2223$ GeV for scan $S_1$ and 
$248-2100$ GeV for $S_2$, {corresponding to $M_2$ in the range
$236 - 1515$ GeV and $169 - 1431$ GeV, respectively}.
Note that for a fixed value of $v_3$, when $Y_{\nu_3}$ is 
sufficiently large the $\widetilde{\nu}_{\tau}$ becomes heavier than 100 GeV,
and these points are not shown in the figure.
{As can also be seen, $Y_{\nu_3}$ acquires larger values in scan $S_2$ than in $S_1$, in agreement with the discussion of Fig.~\ref{S1S2-2D-AvL3-Prefact-MSvL3.png}.}

The values of $Y_{\nu_{3}}$ and $v_3$ used 
in order to obtain a  
$\widetilde{\nu}_{\tau}$ LSP in turn constrain the values of
$Y_{\nu_{1,2}}$ and $v_{1,2}$ producing a correct neutrino physics. 
This is shown in Fig.~\ref{S1S2-NuParams-vs-Yvi.png}, where $\delta m^2$
vs. $Y_{\nu_{i}}$ and $v_i$ is plotted.
%\R{(the other mass difference and mixing angles in Table~\ref{neutralinos-data-used} are also correctly reproduced in these points)}. 
As we can see, we obtain the hierarchy qualitatively discussed in solution 2) of Subsection~\ref{neusneu}, i.e. $Y_{\nu_{3}} < Y_{\nu_{1}} < Y_{\nu_{2}}$, and $v_1 < v_2\lsim v_3$.
{The values of the Yukawas $Y_{\nu_{1,2}}$ in scan $S_2$ are smaller than the corresponding ones in $S_1$ because for these two families the $\nu_R$-Higgsino seesaw contributes significantly to the neutrino masses, and $v_R$ is smaller for scan $S_2$.}
%\R{($v_2$ and $v_3$ are similar but not exactly equal, discuss it?)}.
%\R{ERASE THIS: In particular, we find for $S_1$: $Y_{\nu_{3}}\sim 1.0 - 4.4 \times 10^{-8}$, $Y_{\nu_{1}}\sim 3.5 - 4.5 \times 10^{-7}$, $Y_{\nu_{2}}\sim 7.1 - 8.1 \times 10^{-7}$,
%and $v_1\sim 1.5\times 10^{-5} - 10^{-4}$ GeV, $v_2\sim 8.0\times 10^{-5} - 3.3 \times 10^{-4}$, $v_3\sim 2.0 - 5.3 \times 10^{-4}$, whereas for $S_2$: $Y_{\nu_{3}}\sim 10^{-8} - 2.1 \times 10^{-7}$, $Y_{\nu_{1}}\sim 2.1 - 3.2 \times 10^{-7}$, $Y_{\nu_{2}}\sim 3.8 - 4.7 \times 10^{-7}$, and $v_1\sim 1.7 - 9.7 \times 10^{-5}$ GeV,
%$v_2\sim 7.5\times 10^{-5} - 3.2 \times 10^{-4}$, $v_3\sim 1.6 - 4.7\times 10^{-4}$.}
Concerning the absolute value of neutrino masses, we obtain
$m_{\nu_1}\sim$ 0.002 eV, 
$m_{\nu_2}\sim$ 0.008 eV, and $m_{\nu_3}\sim$ 0.05 eV, 
%Let us finally point out that, consistently with the above discussion
fulfilling the cosmological upper bound on the sum of neutrino masses of $0.12$ eV mentioned in  
Subsection~\ref{choice-of-input-for-scan}.
%Comment here about future CMB constraints?}
The predicted value of the sum of the neutrino masses can be tested in future CMB experiments such as CMB-S4 \cite{Abazajian:2016yjj}. 
{It is also worth noticing here that these hierarchies of neutrino Yukawas and left sneutrino VEVS, give rise to a 
$\widetilde{\nu}_{\mu}$ mass in the range $766-1568$ GeV for scan $S_1$ and 
$466-945$ GeV 
for $S_2$, producing the contributions $a_{\mu}^{\text{SUSY}}\sim 3\times 10^{-10}$ and $\sim 1\times 10^{-10}$, respectively, which are within the SM uncertainty of the muon anomalous magnetic moment as 
mentioned in 
Subsection~\ref{computation-of-likelihoods}.}
%~\ref{choice-of-input-for-scan}}.}

%-----

%Since reproducing neutrino physics is an important asset of the 
%$\mn$, let us discuss first in Subsection~\ref{neutrinos} 
%the constraints that this imposes on the parameter space of the 
%$\widetilde{\nu}_{\tau}$ LSP.
%In Subsection~\ref{pheno}, we will finally analyze the consequences of current and future dilepton displaced-vertex searches on our scenario.

%\clearpage
%%%%%%%%%%%%%%%%%%%%%%%%%%%%%%%%%%%%%%%%%%%%%%%%%%%%%%%%%%%%%%%%%%%%%%%%%%%%%%%%%
%\subsection{Constraints from neutrino physics.}
%\label{neutrinos}

%\R{THINK IF WE HAVE TO INCLUDE THIS kind of discussion:
%We show the correlation between neutrino observables and relevant parameters defined in Eq.~\ref{nu-biai} 
%in Figs.~\ref{fig:S1-aibi-vs-DeltaM.png} and \ref{fig:S1-aibi-over-bmu+btau-vs-theta.png}
%for scans $S_1$  and \ref{fig:S3-aibi-vs-DeltaM.png} and \ref{fig:S3-aibi-over-bmu+btau-vs-theta.png}
%for scans $S_2$ . 
%\begin{eqnarray}
% a_i = Y_{\nu_{ii}} v_u, \quad b_i = Y_{\nu_{ii}} v_d + 3\lambda v_i, \quad c_i = v_i,\quad \frac{1}{M}=\frac{g_1^2}{M_1}+\frac{g_2^2}{M_2}
% \label{nu-biai} 
%\end{eqnarray}
%}

\begin{figure}[t!]
 \centering
\includegraphics[width=\linewidth, height=0.6\textheight]{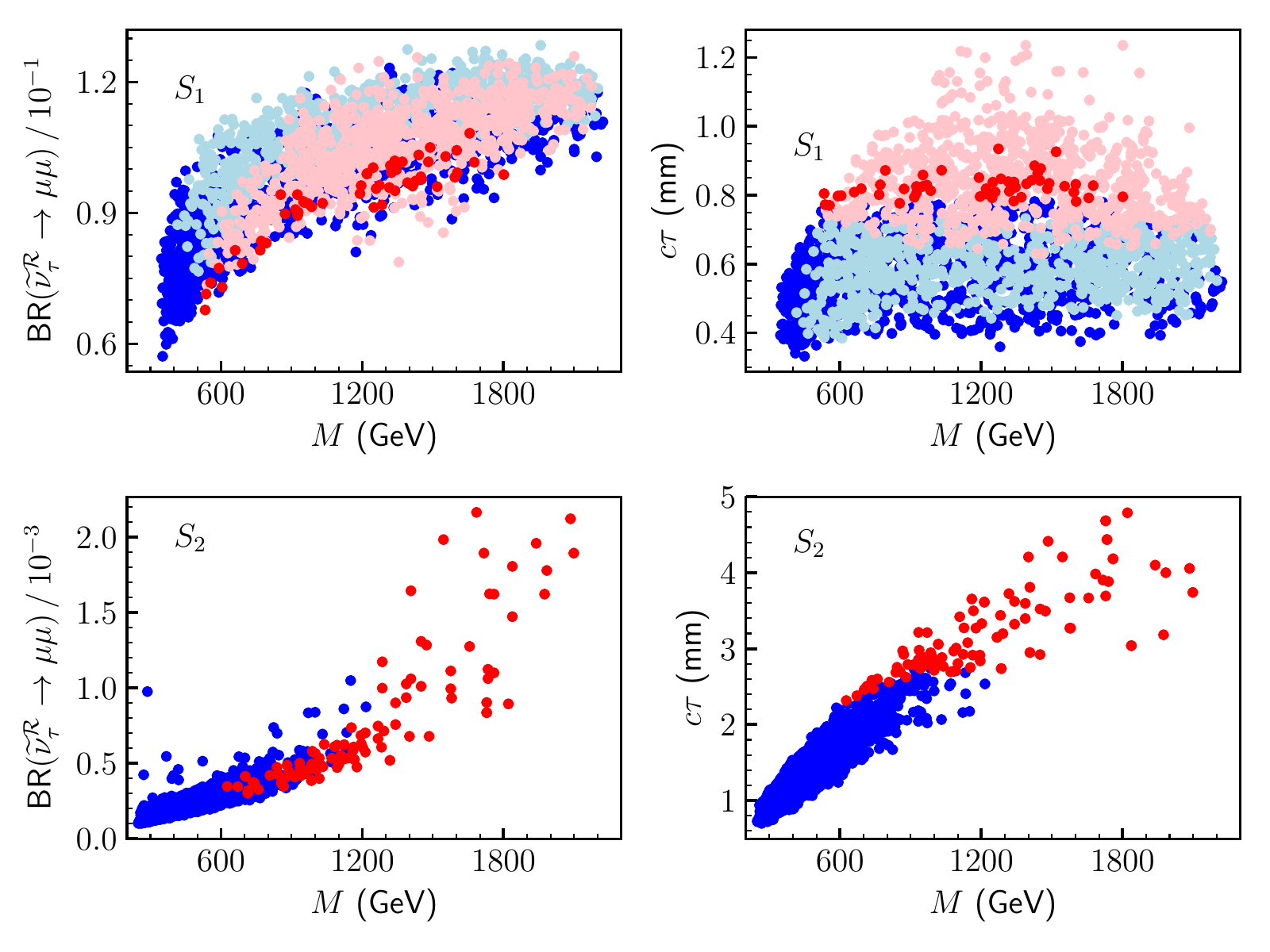}
 \caption{(Left) Branching ratio versus $M$ for the decay of a scalar 
 $\widetilde{\nu}_{\tau}$ LSP with 
 $m_{\widetilde{\nu}_{\tau}}\in (61-100)$ GeV
 into $\mu\mu$ for scan $S_1$ (top) and $S_2$ (bottom).
 (Right) Proper decay length $c\tau$ of the scalar
 $\widetilde{\nu}_{\tau}$ LSP versus $M$
 for scan $S_1$ (top) and $S_2$ (bottom).
 {In all plots,
 the dark-red points indicate that the number of signal events is
 above 3 analyzing the prospects for the 13-TeV search with 
 an integrated luminosity of 300 fb$^{-1}$, combining
 the $\mu\mu$, $e\mu$ and $ee$ channels, and considering also the
 optimization of the trigger requirements discussed in the text. The light-red points in scan $S_1$ although have a number of signal events above 3, are already excluded by the LEP result, as discussed in the text.
 The dark-blue points indicate that the number of signal events is below 3 and therefore inaccessible. The light-blue points in scan $S_1$ have also a number of signal events below 3, and, in addition, are already excluded by the LEP result.} 
 }
 \label{Cut-Re-Brmumu-Mctau.png}
\end{figure}

%\clearpage
%%%%%%%%%%%%%%%%%%%%%%%%%%%%%%%%%%%%%%%%%%%%%%%%%%%%%%%%%%%%%%%%%%%%%%%%%%%%%%%%%
\subsection{Constraints from {accelerator} searches
%on the parameter space for a $\widetilde{\nu}_{\tau}$ LSP.
}
\label{pheno}

Once the neutrino (and sneutrino) physics has determined the relevant regions of the parameter space of the 
$\widetilde{\nu}_{\tau}$ LSP in the $\mn$, we are ready to analyze the reach of the
LHC search.

Given that for each scan the largest neutrino Yukawa is $Y_{\nu_{2}}$,
the most important contribution to the dilepton BRs comes
from the channel $\widetilde{\nu}_{\tau}\rightarrow \tau\mu$.
We also expect
that the BR($\widetilde \nu_\tau \to \mu\mu $) is larger for scan $S_1$
than for $S_2$, as can be checked in
Fig.~\ref{Cut-Re-Brmumu-Mctau.png} (left plots),\footnote{Notice that the partial decay widths into neutrinos for the $S_1$ and $S_2$ cases are similar in size for a given value of $M$, as can be seen from Eq.~\eqref{--sneutrino-decay-width-2nus} and Fig.~\ref{S1S2-NuParams-vs-Yvi.png}. Therefore, a larger partial decay width of the $\widetilde \nu_\tau \to \mu \tau $ channel  for scan $S_1$ implies a larger value of BR($\widetilde\nu^{\mathcal{R}}_{\tau}
\to \mu \mu $), compared with that for scan $S_2$.}
%Figs.~\ref{fig:Cut-Re-Brmumu-Mctau.png}, 
where BR($\widetilde\nu^{\mathcal{R}}_{\tau}
\to \mu\mu $) 
%$\mu\mu$ channel 
is plotted vs. $M$, for the points fulfilling all constraints from neutrino/sneutrino physics (although not shown here, a similar figure is obtained in the case of the pseudoscalar $\widetilde\nu^{\mathcal{I}}_{\tau}$).
The main reason is the smaller (larger) value of $\lambda$ ($\tan\beta$)
for scan $S_1$ with respect to $S_2$, which are crucial parameters in 
Eq.~(\ref{eq:3.22}) for the partial decay width.
Although $\tan\beta$ does not appear explicitly in that equation,
note that 
%$Y_{\tau}$ has a strong dependence on its value,
$Y_{\tau}= ({\sqrt 2 m_{\tau}}/{v})\sqrt{\tan^2\beta + 1}$.
In addition, as shown in Fig.~\ref{S1S2-NuParams-vs-Yvi.png}, the value of $Y_{\nu_{2}}$ is larger for 
scan $S_1$ than for $S_2$, contributing therefore to larger BRs. 
We can also observe in both plots of Fig.~\ref{Cut-Re-Brmumu-Mctau.png} for the BRs that they increase with larger values of $M$. This can be understood from 
Eq.~(\ref{--sneutrino-decay-width-2nus}) 
%and~(\ref{--sneutrino-decay-width-2nus2}) 
showing that larger values of $M$ decrease the decay width to neutrinos.
%\R{(BRs are different for scalars and pseudoscalars, why?).}
In Fig.~\ref{Cut-Re-Brmumu-Mctau.png} (right plots), we show the proper decay length of the $\widetilde\nu^{\mathcal{R}}_{\tau}$
vs. $M$. 
Clearly, this is larger for scan $S_2$ than for $S_1$ because the BRs into charged leptons are smaller in the former case, as discussed before.
Let us finally remember that the lower and upper bounds on $M$ in the figure, have their origin in the analysis of the previous section reproducing neutrino (sneutrino) physics.

It is apparent that in scan $S_2$ for $M$ larger than about 1000 GeV, the points that we find fulfilling all constraints are not uniformly distributed. This happens essentially because the value of $v_R$ is smaller than in scan $S_1$ modifying the relevant contribution of
the $\nu_R$-Higgsino seesaw for the first two families, in such a way that is more difficult to reproduce neutrino physics unless more accurate values of the neutrino Yukawas are input in the computation.
As obtained in Subsection~\ref{neutrinos}, and can be seen in Fig.~\ref{S1S2-NuParams-vs-Yvi.png}, the allowed values of $Y_{\nu_3}$ are larger for $S_2$. This makes more complicated to obtain the correct mixing, producing a tuning in the parameters. 
To obtain these more accurate values, we would have had to run Multinest a much longer time making the task very computer resources demanding. This is not really necessary since it is not going to affect the shape of the figure, and therefore neither the conclusions obtained. 
In addition, let us point out that we could have also modified the values of the parameters used for scan $S_2$ reproducing more easily neutrino physics,
e.g. increasing $v_R$ and modifying accordingly the other parameters to keep the good Higgs physics.

\begin{figure}[t!]
 \centering
\includegraphics[width=\linewidth, height=0.35\textheight]{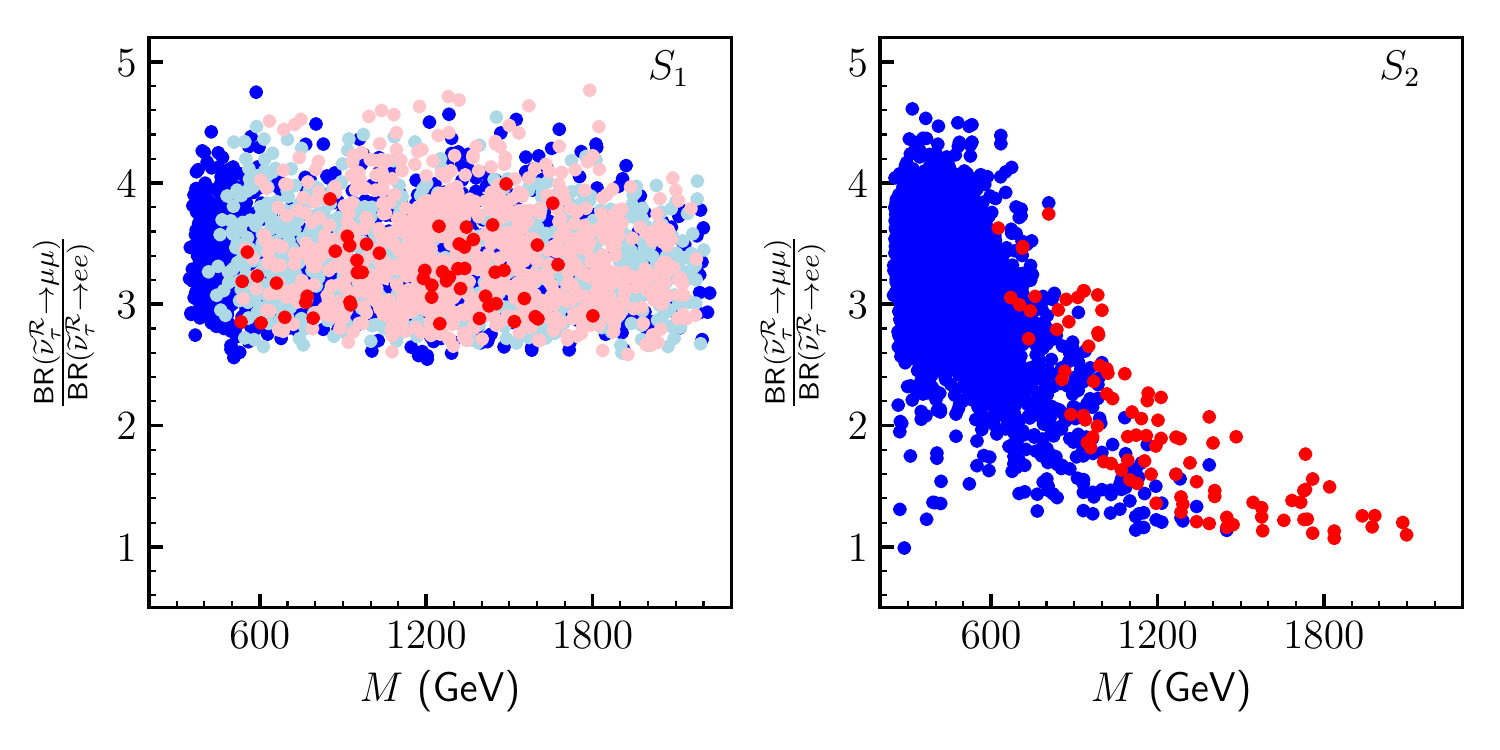}
 \caption{Ratios of the branching fractions of $\widetilde{\nu_\tau}
 \to \mu \mu$ and $\widetilde{\nu_\tau}  \to e e$
 for scan $S_1$ (left) and scan $S_2$ (right).
 {The color code is the same as in Fig.~\ref{Cut-Re-Brmumu-Mctau.png}.}}
 \label{Cut-Re-Brmumu-ee-Ratio}
\end{figure}

In all plots of Fig~\ref{Cut-Re-Brmumu-Mctau.png},
the {(light- and dark-)red} points correspond to regions of the parameter space where the number of signal events is above 3. Note that this only occurs for the 
13-TeV analysis with 
an integrated luminosity of $\mathcal{L}= 300$ fb$^{-1}$.
For the 8-TeV analysis, even considering the optimization of the
trigger requirements, no points have a number of signal events larger than 3. 
%Relevant facts for this result is that the channel 
%$\widetilde \nu_\tau  \to ee$ opens in the 13-TeV case, as %discussed in
%Section~\ref{section2}, and that the value of $Y_{\nu_{1}}$ is significant
%in our scenario,
%increasing therefore the number of events with respect to the 8-TeV case.
%Thus, these regions of the parameter space of the $\mn$ can be probed at the LHC run 3. 
%Our result shows the importance of an optimization not only for the muon trigger but also for the electron trigger. 
{However, we have checked that the light-red points in scan $S_1$ are already excluded by the LEP bound on left sneutrino masses~\cite{Abreu:1999qz,Abreu:2000pi,Achard:2001ek,Heister:2002jc,Abbiendi:2003rn,Abdallah:2003xc}. To carry out this analysis, one can consider e.g. Fig.~6a of Ref.~\cite{Heister:2002jc}, where the cross section upper limit for tau sneutrinos decaying directly to $\ell\ell\tau\tau$ via a dominant $\hat L\hat L\hat e^c$ operator is shown.
Assuming ${\rm BR} = 1$, a lower bound on the sneutrino mass was obtained through the comparison with the MSSM cross section for pair production of tau sneutrinos. 
To recast this result we multiplied this cross section by the factor
$\text{BR}(\widetilde{\nu}_{\tau}^\mathcal{R}\to\tau\mu)\times
\text{BR}(\widetilde{\nu}_{\tau}^\mathcal{I}\to\tau\mu)$ for each of our points.
For an average value of $\text{BR}(\widetilde{\nu}_{\tau}\to\mu\mu)=0.1$ as we can see in Fig~\ref{Cut-Re-Brmumu-Mctau.png}, the cross section must be multiplied then by a factor of $\sim 0.33$, lowering the bound on the sneutrino mass from about 90 GeV in the case of trilinear RPV to about 74 GeV in our case (see Fig.~\ref{Cut-MSvL3-M-Av3.png} below).
This result turns out to be qualitatively different from the one of 
Ref.~\cite{Lara:2018rwv}, where no bound on the sneutrino mass was obtained from recasting the LEP result. This is due to the simplified assumption made in that work that all neutrino Yukawas have the same value and therefore democratic BRs, implying a smaller value for the above factor.}
{On the other hand, 
using Table II of Ref.~\cite{Lara:2018rwv} with the BRs modified appropriately,
we have checked that the lack of constraint on the sneutrino mass from the production of a pair of left staus at LEP obtained in that work, is still valid. We have arrived at the same conclusion for LEP mono-photon search and LHC mono-photon and mono-jet searches, taking also into account the most recent results~\cite{Aaboud:2017dor,Aaboud:2017phn}.} Let us finally remark that
the (light- and dark-)blue points correspond to regions where the number of signal events is below 3, and therefore inaccessible. In addition, we have checked that the light-blue points on top of the dark-blue ones are already excluded by the LEP result.

{Concerning scan $S_2$, we can see in Fig~\ref{Cut-Re-Brmumu-Mctau.png} that the BRs into charged leptons are about two orders of magnitude smaller than for $S_1$, and therefore following the above discussion we have checked that no points are excluded by LEP results in this case.}
Note that although these BRs are smaller, still a significant number of points with signal events above 3 can be obtained when $M$ increases because of the larger value of the decay length, which gives rise to a larger vertex-level efficiency.

%\R{To discuss which regions with red points are these, is sufficient
%Fig.~\ref{Cut-Re-Brmumu-Mctau.png}???}

Figure~\ref{Cut-Re-Brmumu-Mctau.png} also shows that the sensitivity of the dilepton displaced-vertex searches to $\widetilde{\nu}_\tau$ is limited by their small efficiency for $c \tau \lesssim 1$~mm, especially for the $S_1$ case. It is, however, worth noticing that we may even probe such a short lifetime region by optimizing the search strategy for the sub-millimeter displaced vertices, as discussed in Refs.~\cite{Ito:2017dpm, Ito:2018asa}. Our result highly motivates a dedicated work for such an optimization, which we defer to another occasion.

%\R{Add comments here on Fig.~\ref{Cut-Re-Brmumu-ee-Ratio}.}

As discussed in Section~\ref{section2}, the value of $Y_{\nu_{1}}$ is rather large in our scenario, and therefore we expect a sizable branching fraction for the $\widetilde \nu_\tau  \to ee$ channel. In fact, the ratio of the branching fractions for the $\widetilde \nu_\tau  \to ee$ and $\widetilde \nu_\tau  \to \mu \mu$ channels has important implications for our scenario since it reflects the information from the neutrino data via the neutrino Yukawa couplings (see Fig.~\ref{S1S2-NuParams-vs-Yvi.png}). To see this, we plot it against the parameter $M$ in Fig.~\ref{Cut-Re-Brmumu-ee-Ratio}. It is found that for the $S_1$ case, the ratios $R_{\mu/e} \equiv {\rm BR}(\widetilde{\nu}^{\cal R}_\tau \to \mu \mu)/{\rm BR}(\widetilde{\nu}^{\cal R}_\tau \to ee)$ are in the range $3 \lesssim R_{\mu/e} \lesssim 5$, while for the $S_2$ case they are more widely distributed: $1 \lesssim R_{\mu/e} \lesssim 4.6$. 
{This different behaviour can be understood if we realize that for scan $S_1$ 
the second term of $\text{BR}(\widetilde{\nu}_{\tau}\to \mu\mu)$ in Eq.~(\ref{branching}) is negligible with respect to the first one, and the same for the corresponding terms of $\text{BR}(\widetilde{\nu}_{\tau}\to ee)$. Thus, with the approximation in
Eq.~(\ref{eq:3.22}) one gets $R_{\mu/e}\approx (Y_{\nu_{\mu}}/Y_{\nu_{e}})^2$, which using the results for the neutrino Yukawas in Fig.~\ref{S1S2-NuParams-vs-Yvi.png} gives rise to the above range around 3.5. However, for scan $S_2$ the term of $\text{BR}(\widetilde{\nu}_{\tau}\to ee)$ proportional to $\text{BR}(\widetilde{\nu}_{\tau}\to \tau\tau)$ is not negligible
with respect to the one proportional to $\text{BR}(\widetilde{\nu}_{\tau}\to \tau e)$, which is much smaller than in scan $S_1$, due to the contribution of the first term in Eq.~(\ref{eq:3.220}). This implies that the ratio $R_{\mu/e}$ in scan $S_2$ can be smaller than in $S_1$, as can be seen in the figure.}
Now, 
if we particularly focus on the parameter points that can be probed at the 13-TeV LHC, the $S_2$ case predicts $R_{\mu/e} \lesssim 3.6$, and thus we can in principle distinguish this case from the $S_1$ case by measuring this ratio in the future LHC experiments such as the high-luminosity LHC.

\begin{figure}[t!]
 \centering
\includegraphics[width=\linewidth, height=0.35\textheight]{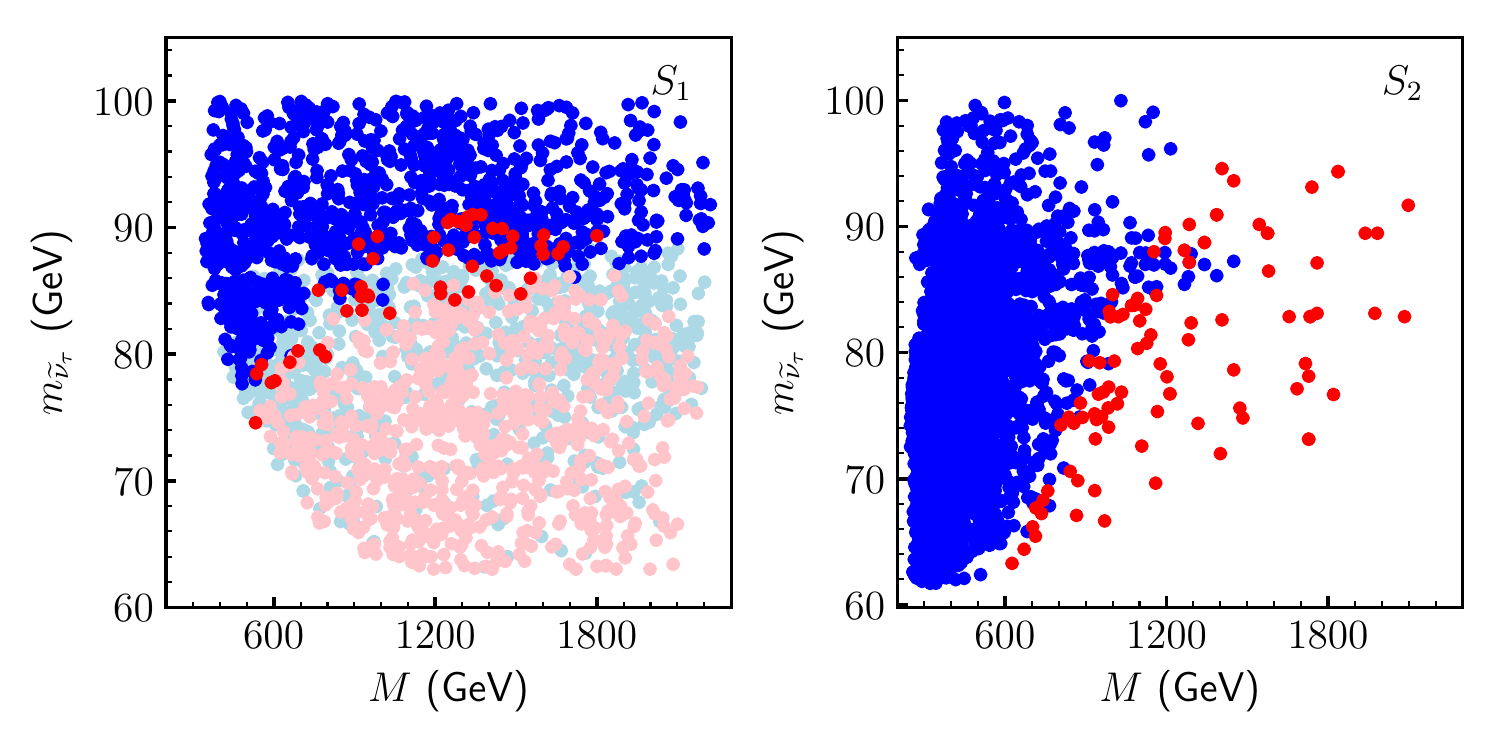}
 \caption{Tau left sneutrino LSP mass versus $M$ 
 %and $A_{\nu_3}$ 
 for scan $S_1$ (left) and scan $S_2$ (right).
{The color code is the same as in Fig.~\ref{Cut-Re-Brmumu-Mctau.png}.}}
 \label{Cut-MSvL3-M-Av3.png}
\end{figure}

Finally, we show in Fig.~\ref{Cut-MSvL3-M-Av3.png} $m_{\widetilde \nu_{\tau}}$ vs. $M$.
For scan $S_1$ (left plot), tau left sneutrino masses in the range $74-91$ GeV
can be probed, 
corresponding to a gaugino mass parameter $M$ in the range
$532-1801$ GeV, {i.e. $M_2\in (363-1228)$ GeV}. Clearly, red points appear in these regions because smaller sneutrino masses produce larger decay lengths.
Since decay lengths are larger for scan $S_2$, the range of sneutrino masses that can be probed is also larger than for $S_1$. In particular, we can see
in the right plot that the range of sneutrino masses is $63-95$ GeV. In this scenario,
$M$
%The corresponding gaugino mass parameter 
is in the range $625-2100$ GeV, correspoding to {$M_2\in (427-1431)$ GeV}.
%\bl{As explained before, the region with points not uniformly distributed can be filled using more computing resources or simply varying the parameters chosen.}
{Let us finally mention that points with sneutrino masses slightly larger than 100 GeV, and with $c\tau > 0.1$ mm, exist, but since they are not constrained by the number of signals events and therefore cannot be probed at the LHC run 3, we do not show them in the figures.}
In any cases, if we actually detect the $\widetilde \nu_\tau$ signal and measure its mass\footnote{As discussed in Ref.~\cite{Lara:2018rwv}, we can in principle measure the mass of $\widetilde \nu_\tau$ by using hadronically decaying tau leptons. } and lifetime in future experiments, we can considerably narrow down the allowed parameter region, which plays an important role in testing the $\mn$.

%\clearpage
%%%%%%%%%%%%%%%%%%%%%%%%%%%%%%%%%%%%%%%%%%%%%%%%%%%%%%%%%%%%%%%%%%%%%%%%%%%%%%%%%%%%%%%%
\section{Conclusions}
\label{Conclusions}

In the framework of the $\mn$, where there is RPV and the several decay BRs of the LSP significantly decrease the signals, there is a lack of experimental bounds on the masses of the sparticles.
To fill this gap in SUSY searches, it is then crucial to analyze the recent experimental results that can lead to limits on sparticle masses in this model, and the prospects for the searches with a higher energy and luminosity.

With this purpose, 
we recast the result of the ATLAS 8-TeV displaced
dilepton search from long-lived particles~\cite{Aad:2015rba}, to obtain the potential limits on the 
parameter space of the tau left sneutrino LSP in the $\mn$ with a mass in the range
$45-100$ GeV.
%from the 8-TeV LHC data.
A crucial point of the analysis, which differentiates the $\mn$ from other SUSY models is that neutrino masses and mixing angles are predicted by the generalized electroweak scale seesaw of the $\mn$ once the parameters of the model are fixed. This is obtained at tree level when three generations of right-handed neutrinos are considered.
Therefore, the sneutrino couplings have to be chosen so that the neutrino 
oscillation data are reproduced, which has important implications for the 
sneutrino decay properties.

%We have {analyzed} the sensitivity of the displaced dilepton searches at
%the LHC to a tau left sneutrino LSP with a mass in the range
%45--100~GeV in the framework of the $\mu\nu$SSM. 
The sneutrino LSP is
produced via the $Z$-boson mediated Drell-Yan process or through the
$W$- and $\gamma/Z$-mediated process accompanied with the production and
decay of the left stau NLSP. Due to the RPV term
present in the $\mu\nu$SSM, the left sneutrino LSP becomes metastable and
eventually decays into the SM leptons. Because of the large
value of the tau Yukawa coupling, a significant
fraction of the sneutrino LSP
decays into a pair of tau leptons or a tau lepton and a light charged
lepton, while the rest decays into a pair of neutrinos. 
A tau sneutrino LSP implies in our scenario that the tau neutrino Yukawa is the smallest coupling, driving neutrino physics to dictate that the muon neutrino Yukawa is the largest of the neutrino Yukawas. As a consequence, 
the most important contribution to the dilepton BRs comes
from the channel $\widetilde{\nu}_{\tau}\rightarrow \tau\mu$.
It is found then that
the decay distance of the left sneutrino tends to be as large as $\gtrsim
1$~mm, which thus can be a good target of displaced vertex searches. 
The strategy that we employed to search for these points was to perform scans of the parameter space of our scenario imposing compatibility with current experimental data on neutrino and Higgs physics, as well as flavor observables.
%We have found that the displaced dilepton search channel is most sensitive to the sneutrino LSP, where at least one of the pair-produced left sneutrinos is required to decay into $\tau \tau$ or $\tau \ell$ with the final-state
%tau leptons decaying leptonically. 

The final result of our analysis for the 8-TeV case is that no points of the parameter space of the $\mn$ can be probed.
This is also true even considering the optimization of the
trigger requirements proposed in Ref.~\cite{Lara:2018rwv}.
%, using a high level trigger that utilizes the tracker information and requires at least one muon with $p_T >$ 24 GeV.
%, we found that no points of the parameter space of the $\mn$ can be probed.
Nevertheless, important regions can be probed at the LHC run 3 with the trigger optimization, as summarized in Fig.~\ref{Cut-MSvL3-M-Av3.png}.
%The relevant fact for this result is that the channel 
%$\widetilde \nu_\tau  \to ee$ opens in the 13-TeV case, where is possible 
%of the trigger requiring at least one electron or muon with $p_T >$ 26 GeV.
%Therefore, the number of events increases with respect to the 8-TeV case.
We in particular emphasize that a trigger optimization for muons has more significant impact on the search ability than that for electrons because of the larger muon neutrino Yukawa coupling in our scenario. 
%such a trigger optimization is of great importance for the search of the tau left sneutrino LSP, and thus we therefore look forward to further analysis in this direction, both from the ATLAS and CMS Collaborations.
Our observation, therefore, suggests that optimizing only the muon trigger already has great benefit. In addition, searching for ``sub-millimeter'' dilepton displaced vertices is also promising. We thus highly motivate both the ATLAS and CMS collaborations to take account of these options seriously. 
 
 If the metastable $\widetilde \nu_\tau$ signature is actually found in the future LHC experiments, we may also measure the mass, lifetime, and decay branching fractions of $\widetilde \nu_\tau$ through the detailed analysis of this signature. We can then include these physical observables into our scan procedure as well in order to further narrow down the allowed parameter space. For instance, we can distinguish the $S_1$ and $S_2$ cases by measuring the ratio ${\rm BR}(\widetilde{\nu}^{\cal R}_\tau \to \mu \mu)/{\rm BR}(\widetilde{\nu}^{\cal R}_\tau \to ee)$ as shown in Fig.~\ref{Cut-Re-Brmumu-ee-Ratio}. We can also restrict the parameter $M$ through the measurements of the mass and decay length of $\widetilde \nu_\tau$, which allows us to infer the gaugino mass scale and thus gives important implications for future high energy colliders.

\begin{acknowledgments}

We would like to thank J. Moreno for his collaboration during the early stages of this work, specially concerning the computing tasks carried out at CESGA.
The work of EK, IL and CM was supported in part by the Spanish Agencia Estatal de Investigaci\'on 
%State Research Agency 
through the grants 
FPA2015-65929-P (MINECO/FEDER, UE), PGC2018-095161-B-I00 and IFT Centro de Excelencia Severo Ochoa SEV-2016-0597.
The work of EK was funded by Fundación La Caixa under
\textquoteleft La Caixa-Severo Ochoa\textquoteright international predoctoral grant.
The work of IL was supported in part by IBS under the project code, IBS-R018-D1
The work of DL was supported by the Argentinian CONICET, and also acknowledges the support of the Spanish grant FPA2015-65929-P (MINECO/FEDER, UE). 
%The authors acknowledge the support of the Spanish Red Consolider MultiDark FPA2017-90566-REDC.
The work of NN was supported in part by the Grant-in-Aid for
Young Scientists B (No.17K14270) and Innovative Areas (No.18H05542).
NN would like to thank the IFT UAM-CSIC for the hospitality of the members of the institute during the Program ``Opportunities at future high energy colliders,'' where this work was finished.
RR acknowledges partial funding/support from the Elusives ITN (Marie Sklodowska-Curie grant agreement No 674896), the  
``SOM Sabor y origen de la Materia'' (FPA 2017-85985-P) and the Spanish MINECO Centro de Excelencia Severo Ochoa del IFIC 
program under grant SEV-2014-0398. EK, IL, CM, DL and RR also acknowledge the support of the Spanish Red Consolider MultiDark FPA2017-90566-REDC.

\end{acknowledgments}

\bibliographystyle{utphys}
\bibliography{munussm_v17}

%\bibliographystyle{utphys}
%\bibliography{munussm_v17}
\end{document}